\documentclass[letterpaper,12pt]{article}
\usepackage{jheppub}
\usepackage{amsmath,amssymb}
\usepackage{bm} %bold math

\def\coeff#1#2{{\textstyle {\frac {#1}{#2}}}}

\def\MO{M_{\scriptscriptstyle \Omega}}
\def\m{{\rm \scriptscriptstyle m}}
\def\e{{\rm \scriptscriptstyle e}}

\def\aBB{\alpha_{\rm\scriptscriptstyle BB}}
\def\cEE{\chi_{\rm\scriptscriptstyle EE}}
\def\cEB{\chi_{\rm\scriptscriptstyle EB}}
\def\cBB{\chi_{\rm\scriptscriptstyle BB}}

\def\GHI{{\rm\scriptscriptstyle d}}
\def\HSR{{\rm\scriptscriptstyle HSR}}
\def\FCRN{{\rm\scriptscriptstyle FCRN}}

\title{ Relativistic magnetohydrodynamics }
\author{Juan Hernandez}
\author{and Pavel Kovtun}
\affiliation{
    Department of Physics and Astronomy, University of Victoria, Victoria, BC, V8P 5C2, Canada
}

\abstract{
We present the equations of relativistic hydrodynamics coupled to dynamical electromagnetic fields, including the effects of polarization, electric fields, and the derivative expansion. We enumerate the transport coefficients at leading order in derivatives, including electrical conductivities, viscosities, and thermodynamic coefficients. We find the constraints on transport coefficients due to the positivity of entropy production, and derive the corresponding Kubo formulas. For the neutral state in a magnetic field, small fluctuations include Alfv\'en waves, magnetosonic waves, and the dissipative modes. For the state with a non-zero dynamical charge density in a magnetic field, plasma oscillations gap out all propagating modes, except for  Alfv\'en-like waves with a quadratic dispersion relation. We relate the transport coefficients in the ``conventional'' magnetohydrodynamics (formulated using Maxwell's equations in matter) to those in the ``dual'' version of magnetohydrodynamics (formulated using the conserved magnetic flux).
}

\begin{document}
\maketitle

\section{Introduction}
In a macroscopic system, near-equilibrium phenomena can often be described by classical hydrodynamics. When the microscopic theory contains weakly coupled $U(1)$ gauge fields, long-range correlations mediated by those fields are possible. Maxwell's equations in matter give an effective description of such correlations in terms of classical gauge fields. These equations  are useful when the coupling between electromagnetic and thermal/mechanical degrees of freedom can be neglected. We would like to understand the effective description of relativistic systems in which macroscopic electromagnetic degrees of freedom are coupled to the macroscopic thermal and mechanical degrees of freedom. This amounts to coupling Maxwell's equations in matter to hydrodynamic equations. When the matter is electrically conducting and electric fields are neglected, such classical effective theory is usually called magneto-hydrodynamics (MHD). 

Our motivation it two-fold. From a fundamental point of view, a number of recent developments in relativistic hydrodynamics have pushed the boundaries of the ``traditional'' theory, as described for example in the classic textbook~\cite{LL6}. These include: a systematic derivative expansion in hydrodynamics~\cite{Baier:2007ix}, an equivalence between hydrodynamics and black hole dynamics~\cite{Bhattacharyya:2008jc}, the manifestation of chiral anomalies in hydrodynamic equations~\cite{Son:2009tf}, the relevance of partition functions~\cite{Banerjee:2012iz, Jensen:2012jh}, elucidation of the role of the entropy current~\cite{Bhattacharyya:2013lha, Bhattacharyya:2014bha}, new insights into relativistic hydrodynamic turbulence~\cite{Fouxon:2009rd}, convergence properties of the hydrodynamic expansion~\cite{Heller:2015dha}, and a classification of hydrodynamic transport coefficients~\cite{Haehl:2015pja}. It is reasonable to expect that the above insights will also lead to an improved understanding of the ``traditional'' MHD. For example, there does not appear to be an agreement in the current literature on such basic question as the number of transport coefficients in MHD. 

From an applied point of view, recent years have seen relativistic hydrodynamics expand from its traditional areas of astrophysical plasmas and hot subnuclear matter into the domain of condensed matter physics. Examples include transport near relativistic quantum critical points~\cite{Hartnoll:2007ih}, in graphene~\cite{Crossno1058, Lucas:2015sya} and in Weyl semi-metals~\cite{Lucas:2016omy}. For conducting matter, MHD is a natural extension of such hydrodynamic models. 

In what follows, we will outline the construction of classical relativistic hydrodynamics with dynamical electromagnetic fields, starting from equilibrium thermodynamics.
In order to write down the hydrodynamic equations, we will assume that the system is locally in thermal equilibrium. We will further assume that the departures from local equilibrium may be implemented through a derivative expansion such that the parameters which characterize the equilibrium (temperature, chemical potential, magnetic field, fluid velocity) vary slowly in space and time. At one-derivative order, transport coefficients such as viscosity and electrical conductivity appear in the constitutive relations. 
We are not aware of previous treatments that list all one-derivative terms in the constitutive relations of magnetohydrodynamics. 

For parity-preserving conducting fluids in magnetic field, we find eleven transport coefficients at one-derivative order. One transport coefficient is thermodynamic, and determines the angular momentum of charged fluid induced by the magnetic field. Three transport coefficients are non-equilibrium and non-dissipative: these are the two Hall viscosities (transverse and longitudinal), and one Hall conductivity. There are also seven non-equilibrium dissipative transport coefficients: two electrical conductivities (transverse and longitudinal), two shear viscosities (transverse and longitudinal), and three bulk viscosities. The constitutive relations for the energy-momentum tensor are given in eqs.~(\ref{eq:TT1}), (\ref{eq:TTF}), and for the  current in eqs.~(\ref{eq:JJ1}), (\ref{eq:JTF}). The dissipative coefficients have to satisfy the inequalities in eq.~(\ref{eq:entropy-constraints}) imposed by the positivity of entropy production, or alternatively by the positivity of the spectral function. As a simple application of the hydrodynamic equations, we study eigenmodes of small oscillations near thermal equilibrium in constant magnetic field. 

We start in Section~\ref{sec:thermodynamics} with a discussion of equilibrium thermodynamics in the presence of external electromagnetic and gravitational fields. In Section~\ref{sec:hydro-1}, we will discuss hydrodynamics, again when electromagnetic and gravitational fields are external. The magnetic fields are taken as ``large'' and electric fields as ``small'' in the sense of the derivative expansion. The smallness of the electric field is due to electric screening. Our procedure will improve on existing studies by taking into account the effects of polarization (magnetic, electric, or both), electric fields, and by enumerating all transport coefficients at leading order in derivatives. In Section~\ref{sec:hydro-2} we discuss hydrodynamics with dynamical electromagnetic fields, as an extension of hydrodynamics with fixed electromagnetic fields. As a simple example, one can study Alfv\'en  and magnetosonic waves in a neutral state (including their damping and polarization), and waves in a dynamically charged (but overall electrically neutral) state. We compare our results with the recent ``dual'' formulation of MHD in Section~\ref{sec:comparison-GHI}, and with some of the previous studies of transport coefficients of relativistic fluids in magnetic field in the Appendix.

\section{Thermodynamics}
\label{sec:thermodynamics}
Let us start with equilibrium thermodynamics. For a system in equilibrium subject to an external non-dynamical gauge field $A_\mu$ and an external non-dynamical metric $g_{\mu\nu}$, we write the logarithm of the partition function $W_s=-i\ln Z$ as 
\begin{equation}
\label{eq:Ws}
  W_s[g,A] = \int\!d^{d+1}x\, \sqrt{-g}\; {\cal F}\,,
\end{equation}
and we will call ${\cal F}$ the free energy density. [Conventions: metric is mostly plus, $\epsilon^{0123}{=}1/\sqrt{-g}$.] For a system with short-range correlations in equilibrium and for external sources $A$ and $g$ which only vary on scales much longer than the correlation length, ${\cal F}$ is a local function of the external sources, and $W_s$ is extensive in the thermodynamic limit. The density ${\cal F}$ may then be written as an expansion in derivatives of the external sources~\cite{Banerjee:2012iz, Jensen:2012jh}. The current $J^\mu$ (defined by varying $W_s$ with respect to the gauge field) and the energy-momentum tensor $T^{\mu\nu}$ (defined by varying $W_s$ with respect to the metric) automatically satisfy
\begin{subequations}
\label{eq:TJ0}
\begin{align}
\label{eq:TJ1}
  & \nabla_{\!\mu} T^{\mu\nu} = F^{\nu\lambda}J_\lambda\,,\\[5pt]
  & \nabla_{\!\mu} J^\mu = 0\,.
\end{align}
\end{subequations}
owing to gauge- and diffeomorphism-invariance of $W_s[g,A]$. The object $W_s[g,A]$ is the generating functional of static (zero frequency) correlation functions of $T^{\mu\nu}$ and $J^\mu$ in equilibrium. Of course, the conservation laws~(\ref{eq:TJ0}) are also true out of equilibrium, being a consequence of gauge- and diffeomorphism-invariance in the microscopic theory. 

Being in equilibrium means that there exists a timelike Killing vector $V$ such that the Lie derivative of the sources with respect to $V$ vanishes.
The equilibrium temperature $T$, velocity $u^\alpha$ and the chemical potential $\mu$ are functions of the Killing vector and the external sources~\cite{Banerjee:2012iz, Jensen:2012jh}
\begin{equation}
\label{eq:Tumu}
  T = \frac{1}{\beta_0 \sqrt{-V^2}}\,,\ \ \ \ 
  u^\mu = \frac{V^\mu}{\sqrt{-V^2}}\,,\ \ \ \ 
  \mu = \frac{ V^\mu A_\mu + \Lambda_V}{\sqrt{-V^2}}\,.
\end{equation}
Here $\beta_0$ is a constant setting the normalization of temperature, and $\Lambda_V$ is a gauge parameter which ensures that $\mu$ is gauge-invariant~\cite{Jensen:2013kka}. The electromagnetic field strength tensor $F_{\mu\nu}=\partial_\mu A_\nu - \partial_\nu A_\mu$ can be decomposed in 3+1 dimensions as
\begin{equation}
\label{eq:FEB}
  F_{\mu\nu} = u_\mu E_\nu - u_\nu E_\mu 
              -\epsilon_{\mu\nu\rho\sigma} u^\rho B^\sigma\,,
\end{equation}
where $E_\mu\equiv F_{\mu\nu}u^\nu$ is the electric field, and $B^\mu \equiv \coeff12 \epsilon^{\mu\nu\alpha\beta} u_\nu F_{\alpha\beta}$ is the magnetic field, satisfying $u{\cdot}E=u{\cdot}B=0$. The decomposition (\ref{eq:FEB}) is just an identity, true for any antisymmetric $F_{\mu\nu}$ and any timelike unit $u^\mu$. Electric and magnetic fields are not independent, but are related by the ``Bianchi identity'' $\epsilon^{\mu\nu\alpha\beta}\nabla_{\!\nu}F_{\alpha\beta}=0$, which in equilibrium becomes
\begin{subequations}
\label{eq:BID-eq}
\begin{align}
  & \nabla{\cdot}B = B{\cdot}a - E{\cdot}\Omega\,,\\[5pt]
  & u_\mu \epsilon^{\mu\nu\rho\sigma}\nabla_{\!\rho} E_\sigma = u_\mu \epsilon^{\mu\nu\rho\sigma}E_\rho a_\sigma\,.
\end{align}
\end{subequations}
Here $\Omega^\mu \equiv \epsilon^{\mu\nu\alpha\beta} u_\nu \nabla_{\!\alpha} u_\beta$ is the vorticity and $a^\mu \equiv u^\lambda \nabla_{\!\lambda}u^\mu$ is the acceleration. In equilibrium, the acceleration is related to temperature by $\partial_\lambda T = - T a_\lambda$. Relations (\ref{eq:BID-eq}) are curved-space versions of the familiar flat-space equilibrium identities ${\bm \nabla}{\cdot}{\bf B}=0$ and ${\bm \nabla}{\times} {\bf E}=0$. 

In order to write down the density ${\cal F}$ in the derivative expansion, we need to specify the derivative counting of the external sources $A$ and $g$. The natural derivative counting for the metric is $g\sim O(1)$ (assuming we are interested in transport phenomena in flat space), while the derivative counting for $A$ depends on the physical system under consideration. 

As an example, consider an insulator, such as a system made out of particles which carry electric/magnetic dipole moments, but no electric charges. In such a system, there is no conserved electric charge, and the above $\mu$ is not a relevant thermodynamic variable. If we are interested in thermodynamics of such a system subject to external electric and magnetic fields, we are free to choose $B\sim O(1)$ and $E\sim O(1)$ in the derivative expansion. The free energy density is then
\begin{equation}
  {\cal F} = p(T, E^2, E{\cdot}B, B^2) + O(\partial)\,.
\end{equation}
The leading-order term is the pressure, whose dependence on $E$ and $B$ encodes the electric, magnetic, and mixed susceptibilities. For the list of $O(\partial)$ contributions to ${\cal F}$, see ref.~\cite{Kovtun:2016lfw}.

As another example, consider a system that has electrically charged degrees of freedom (a conductor), such that $\mu$ gives a non-negligible contribution to thermodynamics. In equilibrium, $\partial_\lambda \mu = E_\lambda - \mu a_\lambda$ is satisfied identically, which suggests that counting $\mu\sim O(1)$ leads to $E\sim O(\partial)$. This is a manifestation of electric screening. The magnetic field, on the other hand, may still be counted as $O(1)$. The counting $B\sim O(1)$ and $E\sim O(\partial)$ is the relevant derivative counting for MHD. The free energy density is then
\begin{equation}
\label{eq:F1}
 {\cal F} = p(T, \mu, B^2) + \sum_{n=1}^{5} M_n(T,\mu,B^2) s_n^{(1)} + O(\partial^2)\,,
\end{equation}
where $s_n^{(1)}$ are $O(\partial)$ gauge- and diffeomorphism-invariants, and the coefficients $M_n$ need to be determined by the microscopic theory, just like the pressure $p$. Following ref.~\cite{Kovtun:2016lfw}, we list the invariants $s_n^{(1)}$ in Table~\ref{tab:T1}. The rows labeled C, P, T indicate the eigenvalue of the invariant under charge conjugation, parity, and time reversal. The last row shows the weight $w$ of the invariant under a local rescaling of the metric: $g_{\mu\nu} \to \tilde g_{\mu\nu} = e^{-2\varphi}g_{\mu\nu}$, and $s_n\to \tilde s_n = e^{w\varphi}s_n$. The invariant $s_3^{(1)}$ does not transform homogeneously under the rescaling, and can not appear in a conformally invariant generating functional. Hence, we expect that in a conformal theory $M_3=0$.
\begin{table}
\begin{center}
\def\arraystretch{1.2}
\setlength\tabcolsep{4pt}
\begin{tabular}{|c|c|c|c|c|c|}
 \hline
 \hline
 $n$ & 1 & 2 & 3 & 4 & 5\\ 
 \hline
 \hline
 $s^{\scriptscriptstyle (1)}_n$
 & $B^\mu \partial_\mu (\frac{B^2}{T^4})$   % 
 & $\epsilon^{\mu\nu\rho\sigma} u_\mu B_\nu \nabla_{\!\rho} B_\sigma$  %
 & $B{\cdot}a$
 & $B{\cdot}\Omega$ 
 & $B{\cdot}E$
 \\
 \hline
  C  & $-$ & $+$ &  $-$ & $-$ & $+$ \\
  \hline
  P  & $-$ & $-$ &  $-$ & $+$ & $-$ \\
  \hline 
  T  & $-$ & $+$ &  $-$ & $+$ & $-$ \\
  \hline
  W  & 3 & 5 & n/a & 3 & 4\\
  \hline
\end{tabular}
\end{center}
\caption{Independent non-zero $O(\partial)$ invariants in equilibrium in 3+1 dimensions.}
\label{tab:T1}
\end{table}
The coefficient $M_5$ is the usual magneto-electric (or electro-magnetic) susceptibility; similarly $M_4$ may be termed magneto-vortical susceptibility.
For the rest of the paper, we will adopt the derivative counting $B\sim O(1)$ and $E\sim O(\partial)$, as is appropriate for MHD.

As an example, consider a parity-invariant theory in magnetic field. The only $O(\partial)$ thermodynamic coefficient is the magneto-vortical susceptibility $\MO\equiv M_4$, which affects $\langle T^{\mu\nu}\rangle$ and $\langle J^\mu\rangle$ when there is non-zero vorticity, and higher-point equilibrium correlation functions of $T^{\mu\nu}$ and $J^\mu$ when there is no vorticity. We define static (zero frequency) correlation functions of $T^{\mu\nu}$ and $J^\mu$ by varying the generating functional (\ref{eq:Ws}) with respect to $g_{\mu\nu}$ and $A_\mu$ in the standard fashion. For example, in flat space at constant temperature $T_0$, constant chemical potential $\mu_0$, and constant magnetic field $B_0$ in the $z$-direction, one finds the following static correlation functions at small momentum
\begin{align}
   \langle T^{tx} J^z \rangle = -k_x k_z \MO\,,\ \ \ \ 
   \langle T^{tx} T^{yz}\rangle = -i B_0 k_z \MO\,.
\end{align}
The first expression may be used to evaluate the magneto-vortical susceptibility $\MO$ in a system that is not subject to magnetic field, and is not rotating.

\section{Hydrodynamics with external electromagnetic fields}
\label{sec:hydro-1}
\subsection{Constitutive relations}
Hydrodynamics is conventionally formulated as an extension of thermodynamics, in the sense that hydrodynamic variables are inherited from the thermodynamic parameters. This is a strong assumption, and we expect the hydrodynamic description only to be valid for $B\ll T^2$, otherwise new non-hydrodynamic degrees of freedom (such as those associated with Landau levels) must be taken into account. Let us start by taking  $E$ and $B$ fields as external and non-dynamical. In hydrodynamics, the thermodynamic variables $T$, $u^\alpha$, and $\mu$ are promoted to time-dependent quantities. Out of equilibrium, they no longer have a microscopic definition, but are merely auxiliary variables used to build the non-equilibrium energy-momentum tensor and the current. The expressions of $T^{\mu\nu}$ and $J^\mu$ in terms of the auxiliary variables $T$, $u^\alpha$, and $\mu$ are called constitutive relations; they contain both thermodynamic contributions (coming from the variation of ${\cal F}$), and non-equilibrium contributions (such as the viscosity). It is worth noting that thermodynamic contributions and non-equilibrium contributions to the constitutive relations may appear at the same order in the derivative expansion. The constitutive relations are then used together with the conservation laws (\ref{eq:TJ0}) to find the energy-momentum tensor and the current. While in thermodynamics Eqs.~(\ref{eq:TJ0}) are mere identities reflecting the symmetries of $W_s$, solving Eqs.~(\ref{eq:TJ0}) in hydrodynamics can be a challenging endeavour leading to rich physics. 

We will write the energy-momentum tensor using the decomposition with respect to the timelike velocity vector $u^\mu$,
\begin{align}
\label{eq:TT1}
   T^{\mu\nu} = {\cal E} u^\mu u^\nu + {\cal P}\Delta^{\mu\nu} + 
    {\cal Q}^\mu u^\nu + {\cal Q}^\nu u^\mu + {\cal T}^{\mu\nu}\,,
\end{align}
where $\Delta^{\mu\nu} \equiv g^{\mu\nu} + u^\mu u^\nu$ is the transverse projector, ${\cal Q}^\mu$ is transverse to $u_\mu$, and  ${\cal T}^{\mu\nu}$ is transverse to $u_\mu$, symmetric, and traceless. Explicitly, the coefficients are ${\cal E}\equiv u_\mu u_\nu T^{\mu\nu}$, ${\cal P}\equiv \coeff{1}{3} \Delta_{\mu\nu}T^{\mu\nu}$, ${\cal Q}_\mu \equiv -\Delta_{\mu\alpha} u_\beta T^{\alpha\beta}$ and ${\cal T}_{\mu\nu}\equiv \coeff12(\Delta_{\mu\alpha}\Delta_{\nu\beta} + \Delta_{\nu\alpha} \Delta_{\mu\beta} - \coeff{2}{3} \Delta_{\mu\nu} \Delta_{\alpha\beta}) T^{\alpha\beta}$. Similarly, we will write the current as
\begin{align}
\label{eq:JJ1}
  J^\mu = {\cal N} u^\mu + {\cal J}^\mu
\end{align}
where the charge density is ${\cal N}\equiv -u_\mu J^\mu$, and the spatial current is ${\cal J}_\mu \equiv \Delta_{\mu\lambda} J^\lambda$.

Using the equilibrium free energy (\ref{eq:F1}), one can isolate $O(1)$ and $O(\partial)$ contributions to the energy-momentum tensor and the current:
\begin{align*}
  & {\cal E} = \epsilon(T,\mu,B^2) + f_{\cal E}\,,\\[5pt]
  & {\cal P} = \Pi(T,\mu,B^2) + f_{\cal P}\,,\\[5pt]
  & {\cal N} = n(T,\mu,B^2) + f_{\cal N}\,,\\[5pt]
  & {\cal T}^{\mu\nu} = \aBB(T,\mu,B^2) \left( B^\mu B^\nu - \coeff13 \Delta^{\mu\nu} B^2 \right)
    +f^{\mu\nu}_{\cal T}\,,
\end{align*}
where $\epsilon = -p+ T(\partial p/\partial T) + \mu (\partial p/\partial\mu)$, $\Pi=p-\coeff23 \aBB B^2$, $n=\partial p/\partial\mu$, and the magnetic susceptibility is $\aBB = 2\partial p/\partial B^2$. The terms $f_{\cal E}$, $f_{\cal P}$, $f_{\cal N}$, $f^{\mu\nu}_{\cal T}$, ${\cal Q}^\mu$, and ${\cal J}^\mu$ are all $O(\partial)$, and contain both equilibrium and non-equilibrium contributions, $f_{\cal E} = \bar f_{\cal E} + f_{\cal E}^\textrm{non-eq.}$ etc, where the bar denotes $O(\partial)$ contributions coming from the variation of $W_s$.

\subsection{Field redefinitions}
Out of equilibrium, the variables $T$, $u^\alpha$, and $\mu$ may be redefined. Such a redefinition is often referred to as a choice of ``frame'', see e.g.\ ref.~\cite{Kovtun:2012rj} for a discussion. Consider changing the hydrodynamic variables to $T'=T+\delta T$, $u'^\alpha = u^\alpha + \delta u^\alpha$, $\mu' = \mu + \delta\mu$, where $\delta T$, $\delta u^\alpha$, and $\delta\mu$ are $O(\partial)$. The same energy-momentum tensor and the current may be expressed either in terms of $T$, $u^\alpha$, $\mu$, or in terms of $T'$, $u'^\alpha$, $\mu'$ (note that $B^2 = B'^2 + O(\partial^2)$). Physical transport coefficients must be derived from $O(\partial)$ quantities which are invariant under such changes of hydrodynamic variables. A direct evaluation shows that the following combinations are invariant under ``frame'' transformations:
\begin{subequations}
\label{eq:fi}
\begin{align}
  & f \equiv f_{\cal P} - \left(\frac{\partial\Pi}{\partial\epsilon}\right)_{\!n} f_{\cal E}
    - \left(\frac{\partial\Pi}{\partial n}\right)_{\!\epsilon} f_{\cal N}\,,\\[5pt]
  & \ell \equiv \frac{B^\alpha}{B} \left( {\cal J}_\alpha - \frac{n}{\epsilon+p}{\cal Q}_\alpha \right)\,,\\[5pt]
  & \ell^\mu_\perp \equiv \mathbb{B}^{\mu\alpha} \left( {\cal J}_\alpha - \frac{n}{\epsilon+p-\aBB B^2}{\cal Q}_\alpha \right) \,,\\[5pt]
  & t^{\mu\nu} \equiv f^{\mu\nu}_{\cal T} - \left( B^\mu B^\nu - \coeff13 \Delta^{\mu\nu}B^2 \right) \left[ \left(\frac{\partial \aBB}{\partial \epsilon}\right)_{\!n} f_{\cal E} + \left(\frac{\partial \aBB}{\partial n}\right)_{\!\epsilon} f_{\cal N}  \right]\,.
\end{align}
\end{subequations}
Here $\mathbb{B}^{\mu\nu} \equiv \Delta^{\mu\nu} - B^\mu B^\nu/B^2$ is the projector onto a plane orthogonal to both $u^\mu$ and $B^\mu$, all thermodynamic derivatives are evaluated at fixed $B^2$, and $B\equiv \sqrt{B^2}$. When the magnetic susceptibility $\aBB$ is $T$- and $\mu$-independent, the stress $f^{\mu\nu}_{\cal T}$ is frame-invariant.

As an example, one can choose $\delta T$ and $\delta\mu$ such that ${\cal E}'=\epsilon(T',\mu',B'^2)$, ${\cal N}'=n(T',\mu',B'^2)$, and further choose $\delta u^\alpha$ such that ${\cal Q}'_\alpha=0$. This corresponds to the Landau-Lifshitz frame~\cite{LL6}. 
The components of energy-momentum tensor and the current take the following form in the Landau-Lifshitz frame:
\begin{subequations}
\label{eq:LL-frame}
\begin{align}
  & {\cal P}' = \Pi(T',\mu',B'^2) + f \,,\\[5pt]
  & {\cal J}'^\mu =  \ell^\mu_\perp + \frac{B'^\mu}{B'} \ell\,,\\[5pt]
  & {\cal T}'^{\mu\nu} = \aBB(T',\mu',B'^2) \left( B'^\mu B'^\nu - \coeff13 \Delta'^{\mu\nu} B'^2 \right)
    +t^{\mu\nu}\,,
\end{align}
\end{subequations}
where the frame invariants are given by eq.~(\ref{eq:fi}). In the Landau-Lifshitz frame, a non-zero value of the pseudoscalar frame-invariant $\ell$ indicates a current flowing along the magnetic field. In a constant external magnetic field such currents arise as consequences of chiral anomalies~\cite{Son:2009tf}; in an inhomogeneous external field, an electric current flowing along the magnetic field can arise without chiral anomalies, owing to a non-zero magnetic susceptibility.

\subsection{Thermodynamic frame}
The energy-momentum tensor and the current derived from the static generating functional $W_s$ correspond to a different frame, termed in~\cite{Jensen:2012jh} the thermodynamic frame. Taking the variation of the free energy (\ref{eq:F1}), one finds the following equilibrium $O(\partial)$ contributions in the thermodynamic frame:
\begin{align}
  & \bar f_{\cal E} = \sum_{n=1}^5 \epsilon_n s_n^{\scriptscriptstyle (1)} \,, \ \ \ \ 
    \bar f_{\cal P} = \sum_{n=1}^5 \pi_n s_n^{\scriptscriptstyle (1)} \,, \ \ \ \ 
    \bar f_{\cal N} = \sum_{n=1}^5 \phi_n s_n^{\scriptscriptstyle (1)} \,, \nonumber\\[5pt]
  & \bar {\cal Q}^\mu = \sum_{n=1}^4 \gamma_n v_n^{{\scriptscriptstyle (1)}\mu} \,, \ \ \ \ 
    \bar {\cal J}^\mu = \sum_{n=1}^4 \delta_n v_n^{{\scriptscriptstyle (1)}\mu} \,, \ \ \ \ 
    \bar f^{\mu\nu}_{\cal T} = \sum_{n=1}^{10} \theta_n t_n^{{\scriptscriptstyle (1)}\mu\nu} \,,
\label{eq:TJeq}
\end{align}
where the bar signifies equilibrium contributions, and the coefficients $\epsilon_n$, $\pi_n$, $\phi_n$, $\gamma_n$, $\delta_n$, $\theta_n$ are all $O(1)$ functions of the five thermodynamic coefficients $M_n(T,\mu,B^2)$ and of the magnetic susceptibility $\aBB=2\partial p/\partial B^2$. The explicit expressions are given in Appendix~\ref{app:thermo}. The one-derivative scalars $s_n^{\scriptscriptstyle (1)}$ are given in Table~\ref{tab:T1}. The one-derivative vectors $v_n^{{\scriptscriptstyle (1)}\mu}$ and tensors $t_n^{{\scriptscriptstyle (1)}\mu\nu}$ are listed in Table~\ref{tab:T2}. The table does not list all $O(\partial)$ vectors and tensors, but only those that appear in the equilibrium ${\cal Q}^\mu$ and ${\cal T}^{\mu\nu}$. 
\begin{table}
\begin{center}
\def\arraystretch{1.2}
\setlength\tabcolsep{4pt}
\begin{tabular}{|c|c|c|c|c|}
 \hline
 \hline
 $n$ & 1 & 2 & 3 & 4 \\ 
 \hline
 \hline
 $v_n^{{\scriptscriptstyle (1)}\mu}$
 & $\epsilon^{\mu\nu\rho\sigma} u_\nu \partial_\sigma B_\rho$   % 
 & $\epsilon^{\mu\nu\rho\sigma} u_\nu B_\rho \partial_\sigma T /T$  %
 & $\epsilon^{\mu\nu\rho\sigma} u_\nu B_\rho \partial_\sigma B^2$
 & $\epsilon^{\mu\nu\rho\sigma} u_\nu E_\rho B_\sigma$ \\
  \hline
\end{tabular}

\bigskip
\begin{tabular}{|c|c|c|c|c|c|c|}
 \hline
 \hline
 $n$ & $1-5$ & 6 & 7 & 8 & 9 & 10\\ 
 \hline
 \hline
 $t_n^{{\scriptscriptstyle (1)}\mu\nu}$
 & $s_n^{\scriptscriptstyle (1)} B^{\langle \mu} B^{\nu\rangle}$   % 
 & $v_1^{{\scriptscriptstyle (1)}} {}^{\langle \mu} B^{\nu\rangle}$
 & $v_2^{{\scriptscriptstyle (1)}} {}^{\langle \mu} B^{\nu\rangle}$
 & $v_3^{{\scriptscriptstyle (1)}} {}^{\langle \mu} B^{\nu\rangle}$ 
 & $v_4^{{\scriptscriptstyle (1)}} {}^{\langle \mu} B^{\nu\rangle}$ 
 & $\Omega^{\langle\mu} B^{\nu\rangle}$ \\
  \hline
\end{tabular}

\end{center}
\caption{Top: Non-zero transverse $O(\partial)$ vectors that appear in the equilibrium energy flux~${\cal Q}^\mu$ and in the equilibrium spatial current ${\cal J}^\mu$. The vector $v_4^{{\scriptscriptstyle (1)}\mu}$ is the Poynting vector. Bottom: Non-zero symmetric transverse traceless $O(\partial)$ tensors that appear in the equilibrium stress ${\cal T}^{\mu\nu}$. For any two transverse vectors $X^\mu$ and $Y^\mu$, the angular brackets stand for $X^{\langle \mu} Y^{\nu\rangle} \equiv X^\mu Y^\nu + X^\nu Y^\mu - \coeff23 \Delta^{\mu\nu} X{\cdot}Y$. }
\label{tab:T2}
\end{table}
The frame invariants (\ref{eq:fi}) then become
\begin{subequations}
\label{eq:fi-2}
\begin{align}
  & f = \sum_{n=1}^5 \Phi_n s_n^{\scriptscriptstyle (1)} + f_\textrm{non-eq.} \,,\ \ \ \ 
  \ell = \sum_{n=1}^5 \Lambda_n s_n^{\scriptscriptstyle (1)} + \ell_\textrm{non-eq.}\,,\\[5pt]
  & \ell^\mu_\perp = \sum_{n=1}^5 \Gamma_{n} v_n^{{\scriptscriptstyle (1)}\mu} + \ell^\mu_{\perp \textrm{non-eq.}}\,,\ \ \ \ 
  t^{\mu\nu} = \sum_{n=1}^{10} \Theta_n t_n^{{\scriptscriptstyle (1)}\mu\nu} + t^{\mu\nu}_\textrm{non-eq.}
\end{align}
\end{subequations}
In the vector invariant, we have defined $v_5^{{\scriptscriptstyle (1)}\mu} \equiv s_2^{\scriptscriptstyle (1)} B^\mu$.
The subscript ``non-eq'' denotes non-equilibrium contributions which by definition vanish in equilibrium. The functions $\Phi_n(T,\mu,B^2)$, $\Lambda_n(T,\mu,B^2)$, $\Gamma_n(T,\mu,B^2)$, $\Theta_n(T,\mu,B^2)$ are non-dissipative thermodynamic transport coefficients. Explicitly,
\begin{align*}
  & \Phi_n = \pi_n - \epsilon_n \left(\frac{\partial\Pi}{\partial\epsilon}\right)_{\!n} - \phi_n \left(\frac{\partial\Pi}{\partial n}\right)_{\!\epsilon} \,, \ \ \ \ 
   \Lambda_{n\neq2} = 0\,,\ \ \ \ 
   \Lambda_2 = \frac{1}{B} \big(\delta_1 - \frac{n}{\epsilon+p} \gamma_1 \big) \,,\\[5pt]
  & \Gamma_{n\leqslant4} = \delta_n - \frac{n}{\epsilon{+}p{-}\aBB B^2} \gamma_n \,,\ \ \ \ \Gamma_5 = -\frac{1}{B^2} \left( \delta_1 - \frac{n}{\epsilon{+}p{-}\aBB B^2} \gamma_1 \right) \,,\\[5pt]
  & \Theta_{n\leqslant 5} = \theta_n - \coeff12 \epsilon_n \left(\frac{\partial\aBB}{\partial\epsilon}\right)_{\!n} - \coeff12  \phi_n \left(\frac{\partial\aBB}{\partial n}\right)_{\!\epsilon}  \,,\ \ \ \ \ 
   \Theta_{n\geqslant 6} = \theta_n \,.
\end{align*}
We see that the constitutive relations for energy-momentum tensor and the current contain twenty-one thermodynamic transport coefficients $\Phi_n$, $\Lambda_2$, $\Gamma_n$, $\Theta_n$. These twenty-one coefficients are not independent, but can all be expressed in terms of only five parameters $M_n$ of the equilibrium generating functional. 

Let us now write down the constitutive relations in the thermodynamic frame that is a natural generalization of the Landau-Lifshitz frame. We will define the thermodynamic frame (primed variables) by redefinitions of $T$, $\mu$, and $u^\alpha$ that give 
\begin{subequations}
\label{eq:CRTF}
\begin{align}
 & {\cal E}' = \epsilon(T',\mu',B'^2) + \bar f_{\cal E}\,,\\[5pt]
 & {\cal N}' = n(T',\mu',B'^2) + \bar f_{\cal N}\,,\\[5pt]
 & {\cal Q}'_\alpha = \bar{\cal Q}_\alpha\,.
\end{align}
In other words, in this thermodynamic frame the coefficients ${\cal E}$, ${\cal N}$, and ${\cal Q}_\alpha$ in the decompositions (\ref{eq:TT1}), (\ref{eq:JJ1}) take their equilibrium values, derived from the equilibrium generating functional~$W_s$. The other coefficients take the following form in the thermodynamic frame:
\begin{align}
 & {\cal P}' = \Pi(T',\mu',B'^2) + \bar f_{\cal P} + f_\textrm{non-eq.}\,,\\[5pt]
 & {\cal J}'^\mu = \bar {\cal J}^\mu + \ell^\mu_{\perp\textrm{non-eq.}} + \frac{B'^\mu}{B'} \ell_\textrm{non-eq.}\,,\\[5pt]
 & {\cal T}'^{\mu\nu} = \aBB(T',\mu',B'^2) \left( B'^\mu B'^\nu - \coeff13 \Delta'^{\mu\nu} B'^2 \right) + \bar f_{\cal T}^{\mu\nu} + t^{\mu\nu}_\textrm{non-eq.}\,.
\end{align}
\end{subequations}

\subsection{Non-equilibrium contributions}
With the equilibrium contributions out of the way, the next task is to find the non-equilibrium terms in the constitutive relations (\ref{eq:fi-2}). This amounts to finding one-derivative scalars, vectors (orthogonal both to $B_\mu$ and to $u_\mu$), and transverse traceless symmetric tensors that vanish in equilibrium. Note that non-equilibrium contributions (those that vanish in equilibrium) are not the same as dissipative contributions (those that contribute to hydrodynamic entropy production). Every dissipative contribution is non-equilibrium, but not every non-equilibrium contribution is dissipative.

\begin{table}
\begin{center}
\def\arraystretch{1.2}
\setlength\tabcolsep{4pt}
\begin{tabular}{|c|c|c|c|c|c|c|}
 \hline
 \hline
 $n$ & 1 & 2 & 3 & 4 & 5 & 6\\ 
 \hline
 \hline
 $s_{n\textrm{ non-eq.}}^{{\scriptscriptstyle (1)}}$
 & $u^\lambda \partial_\lambda T$
 & $u^\lambda \partial_\lambda \mu$  %
 & $\nabla{\cdot}u$
 & $b^\mu b^\nu \nabla_{\!\mu} u_\nu$ 
 & $b^\lambda E_\lambda - T b^\lambda \partial_\lambda(\mu/T)$ 
 & $b^\lambda a_\lambda + b^\lambda \partial_\lambda T /T$ \\
  \hline
 P & $+$ & $+$ & $+$ & $+$ & $-$ & $-$\\
 \hline 
\end{tabular}

\bigskip
\begin{tabular}{|c|c|c|c|}
 \hline
 \hline
 $n$ & 1 & 2 & 3 \\ 
 \hline
 \hline
 $v_{n\textrm{ non-eq.}}^{{\scriptscriptstyle (1)}\mu}$
 & $E^\mu - T \Delta^{\mu\nu} \partial_\nu(\mu/T)$   % 
 & $a^\mu + \Delta^{\mu\nu} \partial_\nu T/T$  %
 & $\sigma^{\mu\nu} b_\nu$ \\
  \hline
 P & $-$ & $-$ & $+$\\
 \hline
\end{tabular}
\end{center}
\caption{
Non-equilibrium scalars and transverse non-equilibrium vectors at $O(\partial)$, written in terms of $b^\mu \equiv B^\mu /B$. In addition to the vectors listed in the table, there are corresponding transverse non-equilibrium vectors $\tilde v^{{\scriptscriptstyle (1)}\mu}_\textrm{non-eq.} \equiv \epsilon^{\mu\nu\rho\sigma}u_\nu b_\rho v^{\scriptscriptstyle (1)}_{\textrm{non-eq.}\,\sigma}$. The table also shows the parity of non-equilibrium scalars and vectors. Under time-reversal, the scalars $s_{n\textrm{ non-eq.}}^{{\scriptscriptstyle (1)}}$ are T-odd, the vectors $v_{n\textrm{ non-eq.}}^{{\scriptscriptstyle (1)}\mu}$ are T-even, and the vectors $\tilde v_{n\textrm{ non-eq.}}^{{\scriptscriptstyle (1)}\mu}$ are T-odd.
}
\label{tab:T3}
\end{table}

The six independent non-equilibrium one-derivative scalars are given in Table~\ref{tab:T3}. 
The scalar $u^\lambda \partial_\lambda B^2$ is not independent as a consequence of the electromagnetic Bianchi identity, and can be expressed as a combination of $\nabla{\cdot}u$ and $B^\mu B^\nu \nabla_{\!\mu} u_\nu$. Three scalar equations of motion $\nabla_{\!\mu}J^\mu = 0$, $u_\nu \nabla_{\!\mu} T^{\mu\nu} + E_\mu J^\mu=0$, and $B_\nu \nabla_{\!\mu} T^{\mu\nu} + (E{\cdot}B) (u{\cdot}J)=0$ taken at zeroth order provide three relations among the scalars. We choose to eliminate $s^{\scriptscriptstyle (1)}_{1 \textrm{ non-eq.}}$, $s^{\scriptscriptstyle (1)}_{2 \textrm{ non-eq.}}$, and $s^{\scriptscriptstyle (1)}_{6 \textrm{ non-eq.}}$
and write the scalar and pseudo-scalar constitutive relations as
\begin{align*}
  & f_\textrm{non-eq.} = c_{1} s^{\scriptscriptstyle (1)}_{3 \textrm{ non-eq.}} + c_{2} s^{\scriptscriptstyle (1)}_{4 \textrm{ non-eq.}} + c_{3} s^{\scriptscriptstyle (1)}_{5 \textrm{ non-eq.}} \,,\\[5pt]
  & \ell_\textrm{non-eq.} = c_{4} s^{\scriptscriptstyle (1)}_{3 \textrm{ non-eq.}} + c_{5} s^{\scriptscriptstyle (1)}_{4 \textrm{ non-eq.}} + c_{6} s^{\scriptscriptstyle (1)}_{5 \textrm{ non-eq.}} \,,
\end{align*}
with some undetermined transport coefficients $c_n$.

The independent non-equilibrium transverse one-derivative vectors are given in Table~\ref{tab:T3}, where the shear tensor is $\sigma^{\mu\nu} \equiv \Delta^{\mu\alpha} \Delta^{\nu\beta}(\nabla_\alpha u_\beta + \nabla_\beta u_\alpha -\coeff23 \Delta_{\alpha\beta} \nabla{\cdot}u)$. 
We use the vector equation of motion (\ref{eq:TJ1}) projected with $\mathbb{B}^{\mu\nu}$ at zeroth order to eliminate one of the vectors,%
\footnote{
Namely, using the equation of motion (\ref{eq:TJ1}) with the constitutive relations for $T^{\mu\nu}$ and $J^\mu$ derived from the generating functional $W=\int\!\sqrt{-g}\, p(T,\mu,B^2)+O(\partial)$. The relation among the vectors that one finds is $v_{2\textrm{ non-eq.}}^{{\scriptscriptstyle (1)}\mu} = v_{1\textrm{ non-eq.}}^{{\scriptscriptstyle (1)}\mu} n/(\epsilon+p) + O(\partial^2)$. 
}
and write the vector constitutive relation as
$$
  \ell^\mu_{\perp\textrm{non-eq.}} = c_7\, \mathbb{B}^{\mu}_{\ \nu}\, v_{1\textrm{ non-eq.}}^{{\scriptscriptstyle (1)}\nu}  + c_8\, \mathbb{B}^{\mu}_{\ \nu}\, v_{3\textrm{ non-eq.}}^{{\scriptscriptstyle (1)}\nu} + c_{9}\, \tilde v_{1\textrm{ non-eq.}}^{{\scriptscriptstyle (1)}\mu} + c_{10}\, \tilde v_{3\textrm{ non-eq.}}^{{\scriptscriptstyle (1)}\mu}  \,,
$$
The tilded vectors are defined as $\tilde v^\mu \equiv \epsilon^{\mu\nu\rho\sigma}u_\nu B_\rho v_\sigma/B$.

There is a number of symmetric transverse traceless non-equilibrium one-derivative tensors besides the shear tensor $\sigma^{\mu\nu}$. One such tensor is
\begin{equation}
\label{eq:sigma-tilde}
  \tilde\sigma^{\mu\nu} \equiv \frac{1}{2B} \left( \epsilon^{\mu\lambda\alpha\beta} u_\lambda B_\alpha \sigma_\beta^{\ \ \nu} + \epsilon^{\nu\lambda\alpha\beta} u_\lambda B_\alpha \sigma_\beta^{\ \ \mu}\right)\,.
\end{equation}
Other tensors can be formed by $B^{\langle\mu} B^{\nu\rangle} s^{\scriptscriptstyle (1)}_{n \textrm{ non-eq.}}$, or by symmetrizing $B^\mu$ with a transverse non-equilibrium vector. Again, we eliminate three scalars and one vector by the zeroth order equations of motion and  write the tensor constitutive relation in terms of $b^\mu \equiv B^\mu/B$ as 
\begin{align*}
  t^{\mu\nu}_\textrm{non-eq.} & = c_{11} \sigma^{\mu\nu} +  b^{\langle\mu} b^{\nu\rangle} \left( c_{12} s^{\scriptscriptstyle (1)}_{3 \textrm{ non-eq.}} + c_{13} s^{\scriptscriptstyle (1)}_{4 \textrm{ non-eq.}} + c_{14} s^{\scriptscriptstyle (1)}_{5 \textrm{ non-eq.}}\right) \\[5pt]
  & + c_{15}b^{\langle\mu} v_{1\textrm{ non-eq.}}^{{\scriptscriptstyle (1)}\nu \rangle} + c_{16} b^{\langle\mu} v_{3\textrm{ non-eq.}}^{{\scriptscriptstyle (1)}\nu \rangle} + c_{17} b^{\langle\mu} \tilde v_{1\textrm{ non-eq.}}^{{\scriptscriptstyle (1)}\nu \rangle} + c_{18} b^{\langle\mu} \tilde v_{3\textrm{ non-eq.}}^{{\scriptscriptstyle (1)}\nu \rangle} + c_{19}\, \tilde\sigma^{\mu\nu}\,,
\end{align*}
with some undetermined transport coefficients $c_n$. Thus there are five equilibrium functions $M_n(T,\mu,B^2)$, and nineteen non-equilibrium functions $c_n(T,\mu,B^2)$ that determine one-derivative contributions to the energy-momentum tensor and the current in strong magnetic field.
If the microscopic system is parity-invariant, all thermodynamic coefficients $M_n$ vanish except for $M_4$. In addition, the dynamical coefficients $c_3$, $c_4$, $c_5$, $c_8$, $c_{10}$, $c_{14}$, $c_{15}$, $c_{17}$ must vanish by parity invariance. Thus a conducting parity-invariant system in magnetic field has one thermodynamic coefficient $M_4$, three ``electrical conductivities'' $c_6$, $c_7$, and $c_9$, and eight ``viscosities'' $c_1$, $c_2$, $c_{11}$, $c_{12}$, $c_{13}$, $c_{16}$, $c_{18}$, and~$c_{19}$. We will see later that the Onsager relations impose a relation between $c_2$, $c_{12}$, and $c_{13}$, plus four more relations among the parity-violating coefficients. This leaves eleven transport coefficients (one thermodynamic and ten non-equilibrium) for a conducting parity-invariant system in magnetic field in 3+1 dimensions. In a conformal theory, the tracelessness condition%
\footnote{
  In a conformal theory subject to external fields $g_{\mu\nu}$ and $A_\mu$, the trace of the energy-momentum tensor receives an anomalous contribution $T^\mu_{\ \mu} = \kappa F^2 + O(\partial^4)$, where $\kappa$ is a theory-dependent constant that counts the number of charged degrees of freedom, and the terms $O(\partial^4)$ are due to curvature invariants. It was shown in ref.~\cite{Eling:2013bj} that the conformal anomaly may be captured by a certain local term in the hydrostatic generating functional, which for our purposes amounts to a term in $p(T,\mu,B^2)$ proportional to $\kappa$.
}
will in addition impose $c_1 = c_2 = 0$.

The constitutive relations may be simplified further if we note that the shear tensor can be decomposed with respect to the magnetic field as
\begin{equation}
  \sigma^{\mu\nu} = \sigma^{\mu\nu}_\perp + \left( b^\mu \Sigma^\nu + b^\nu \Sigma^\mu \right) + \coeff12 {b^{\langle \mu} b^{\nu\rangle}} \left( 3 S_4 - S_3 \right)\,.
\end{equation}
Here $\sigma^{\mu\nu}_\perp \equiv \coeff12 \left(\mathbb{B}^{\mu\alpha} \mathbb{B}^{\nu\beta} + \mathbb{B}^{\nu\alpha} \mathbb{B}^{\mu\beta} - \mathbb{B}^{\mu\nu} \mathbb{B}^{\alpha\beta}\right) \sigma_{\alpha\beta}$ is traceless, $\Sigma^\mu \equiv \mathbb{B}^{\mu\lambda}\sigma_{\lambda\rho}b^\rho$, and both are orthogonal to the magnetic field $B_\mu$. The scalars are $S_3\equiv \nabla{\cdot}u$ and $S_4 \equiv b^\mu b^\nu \nabla_{\!\mu}u_\nu$. The tensor (\ref{eq:sigma-tilde}) then becomes
\begin{equation}
  \tilde\sigma^{\mu\nu} = \tilde\sigma^{\mu\nu}_\perp + \coeff12 \left( b^\mu \tilde\Sigma^\nu + b^\nu \tilde\Sigma^\mu \right)\,,
\end{equation}
where $\tilde\sigma^{\mu\nu}_\perp$ is transverse to both $u_\mu$ and $B_\mu$, symmetric, and traceless.

For completeness, let us summarize the constitutive relations for a parity-invariant theory in the thermodynamic frame. Defining $\MO\equiv M_4$, the energy-momentum tensor is given by eq.~(\ref{eq:TT1}) with the following coefficients:
\begin{subequations}
\label{eq:TTF}
\begin{align}
  {\cal E} & =  -p + T\, p_{,T} + \mu\, p_{,\mu}
  +\left(T {\MO}_{,T} + \mu {\MO}_{,\mu} -2 \MO \right) B{\cdot}\Omega\,,\\[5pt]
  {\cal P} & = p - \coeff43\, p_{,B^2}B^2
  -\coeff13 (\MO + 4{\MO}_{,B^2}B^2)B{\cdot}\Omega
  -\zeta_1 \nabla{\cdot}u - \zeta_2 {b^\mu b^\nu}\nabla_{\!\mu} u_\nu\,,\\[5pt]
  {\cal Q}^\mu & = - \MO \epsilon^{\mu\nu\rho\sigma} u_\nu \partial_\sigma B_\rho
  +(2 \MO - T {\MO}_{,T}- \mu {\MO}_{,\mu}) \epsilon^{\mu\nu\rho\sigma} u_\nu B_\rho \partial_\sigma T /T
  \nonumber\\[5pt]
  & \ \ \ \,
  - {\MO}_{,B^2} \epsilon^{\mu\nu\rho\sigma} u_\nu B_\rho \partial_\sigma B^2
  +(-2p_{,B^2} + {\MO}_{,\mu} - 2 {\MO}_{,B^2} B{\cdot}\Omega) \epsilon^{\mu\nu\rho\sigma} u_\nu E_\rho B_\sigma
  \nonumber\\[5pt]
  & \ \ \ \,
  + \MO \epsilon^{\mu\nu\rho\sigma}\Omega_\nu E_\rho u_\sigma\,, \\[5pt]
  {\cal T}^{\mu\nu} & = 2p_{,B^2} \left( B^\mu B^\nu -\coeff13 \Delta^{\mu\nu}B^2\right)
  + {\MO}_{,B^2} B^{\langle\mu} B^{\nu\rangle} B{\cdot}\Omega
  + \MO B^{\langle\mu}\Omega^{\nu\rangle} \nonumber \\[5pt]
  & \ \ \ \, - \eta_\perp \sigma^{\mu\nu}_\perp
  -\eta_\parallel (b^\mu \Sigma^\nu + b^\nu \Sigma^\mu) 
  - b^{\langle \mu}b^{\nu\rangle} \left(\eta_1 \nabla{\cdot}u + \eta_2 b^\alpha b^\beta \nabla_\alpha u_\beta \right)\nonumber\\[5pt]
  & \ \ \ \,  
  - \tilde\eta_\perp \tilde\sigma^{\mu\nu}_\perp 
  - \tilde\eta_\parallel (b^\mu \tilde\Sigma^\nu + b^\nu \tilde\Sigma^\mu)\,,
\end{align}
\end{subequations}
and the current is given by eq.~(\ref{eq:JJ1}) with the following coefficients:
\begin{subequations}
\label{eq:JTF}
\begin{align}
  {\cal N} & = p_{,\mu} + {\MO}_{,\mu}B{\cdot}\Omega - m {\cdot} \Omega\,,\\[5pt]
  {\cal J}^\mu & = \epsilon^{\mu\nu\rho\sigma} u_\nu \nabla_{\!\rho} m_\sigma
  +\epsilon^{\mu\nu\rho\sigma} u_\nu a_\rho m_\sigma 
  + \left( \sigma_\perp \mathbb{B}^{\mu\nu} + \sigma_\parallel \frac{B^\mu B^\nu}{B^2}\right) V_\nu +  \tilde\sigma\, \tilde V^\mu  \,.
\end{align}
\end{subequations}
The current is written in terms of the magnetic polarization vector
\begin{equation}
\label{eq:m-vector}
  m^\mu = \left( 2\,p_{,B^2} + 2 {\MO}_{,B^2}B{\cdot}\Omega \right)B^\mu + \MO \Omega^\mu\,,
\end{equation}
while the electric polarization vector vanishes at leading order in a parity-invariant system.
The comma subscript denotes the derivative with respect to the argument that follows. Note that we are keeping $O(\partial^2)$ thermodynamic terms in the constitutive relations (coming from the variation of $M_4 s^{\scriptscriptstyle (1)}_4$) that are needed to ensure that the conservation laws (\ref{eq:TJ0}) are satisfied identically for time-independent background fields. In writing down the constitutive relations (\ref{eq:TTF}), (\ref{eq:JTF}), we have relabeled the non-equilibrium transport coefficients as $\zeta_1 \equiv -c_1$, $\zeta_2\equiv - c_2$, $\sigma_\parallel \equiv c_6$, $\sigma_\perp \equiv c_7$, $\tilde\sigma\equiv c_9$, $\eta_\perp \equiv -c_{11}$, $\eta_\parallel \equiv -c_{11} - c_{16}$, $\eta_1\equiv - c_{12} +\frac12 c_{11} +\frac23 c_{16}$, $\eta_2 \equiv - c_{13} - \frac32 c_{11} - 2c_{16}$, $\tilde\eta_\parallel \equiv - c_{18} - \frac12 c_{19}$, $\tilde\eta_\perp \equiv -c_{19}$, and defined $V^\mu \equiv E^\mu - T \Delta^{\mu\nu} \partial_\nu(\mu/T)$.
The coefficients $\sigma_\perp$, $\sigma_\parallel$ are the transverse and longitudinal conductivities, and $\eta_\perp$, $\eta_\parallel$ are the transverse and longitudinal shear viscosities. The coefficients $\zeta_1$, $\zeta_2$, $\eta_1$ and $\eta_2$ may all be called ``bulk viscosities'', of which only three are independent due to the Onsager relation. The coefficients $\tilde\eta_\perp$, $\tilde\eta_\parallel$ are the two Hall viscosities, and $\tilde\sigma$ is the Hall conductivity.%
\footnote{
  The actual Hall conductivity, measured as a response to external electric field, must be obtained after the hydrodynamic equations with the constitutive relations (\ref{eq:TTF}), (\ref{eq:JTF}) have been solved. Doing so in a state with constant charge density $n_0$ and magnetic field $B_0$ gives the Hall conductivity $n_0/B_0$, as expected from elementary considerations of boosting the state in the plane transverse to ${\bf B}_0$. See eq.~(\ref{eq:Kubo-r1c}) below.
}

When the external electromagnetic field vanishes, the system becomes isotropic, and we expect to recover the constitutive relations of the standard isotropic hydrodynamics, with shear viscosity $\eta$, bulk viscosity $\zeta$, and electrical conductivity $\sigma$. Thus as $B\to 0$ we expect $\eta_\perp = \eta_\parallel = -2\eta_1 = \frac23 \eta_2 = \eta$, $\tilde\eta_\perp = \tilde\eta_\parallel = 0$, $\zeta_1 = \zeta$, $\zeta_2 = 0$, $\sigma_\perp = \sigma_\parallel = \sigma$, $\tilde\sigma=0$.

\subsection{Eigenmodes}
\label{sec:eigenmodes-1}
As a simple application of the hydrodynamic equations (\ref{eq:TJ0}) together with the constitutive relations (\ref{eq:TTF}), (\ref{eq:JTF}), one can study the eigenmodes of small oscillations about the thermal equilibrium state. We set the external sources to zero, and linearize the hydrodynamic equations near the flat-space equilibrium state with constant $T=T_0$, $\mu=\mu_0$, $u^\alpha=(1,{\bf 0})$, and $B^\alpha=(0,0,0,B_0)$. Taking the fluctuating hydrodynamic variables proportional to $\exp(-i\omega t+ i{\bf k}{\cdot}{\bf x})$, the source-free system admits five eigenmodes, two gapped ($\omega({\bf k}{\to}0)\neq0$), and three gapless ($\omega({\bf k}{\to}0)=0$). The frequencies of the gapped eigenmodes are
\begin{align}
\label{eq:omega-cyclo}
  \omega = \pm\frac{B_0 n_0}{w_0} - \frac{i B_0^2}{w_0} \left( \sigma_\perp \pm i\tilde\sigma\right)
  -i D_c k^2\,,
\end{align}
where $w_0 \equiv \epsilon_0 + p_0$ is the equilibrium enthalpy density, and we have taken $\aBB B_0^2 \ll w_0$, ${\MO}_{,\mu} B_0^2 \ll w_0$ in the hydrodynamic regime $B_0\ll T_0^2$. As the imaginary part of the eigenfrequency must be negative for stability, this implies $\sigma_\perp >0$. The mode has a circular polarization (at $k=0$), with $\delta u_x$ and $\delta u_y$ oscillating with a $\pi/2$ phase difference. The analogous mode in 2+1 dimensional hydrodynamics was christened the hydrodynamic cyclotron mode in ref.~\cite{Hartnoll:2007ih}, which also explored its implications for transport near two-dimensional quantum critical points. 

For momenta ${\bf k}\parallel {\bf B}_0$, the three gapless eigenmodes are the two sound waves, and one diffusive mode. The eigenfrequencies in the small momentum limit are
\begin{subequations}
\label{eq:omega-par}
\begin{align}
  & \omega = \pm k v_s - i\frac{\Gamma_{s,\parallel}}{2} k^2\,,  \\
  & \omega = -iD_\parallel k^2\,,
\end{align}
\end{subequations}
where $v_s$ is the speed of sound. 
As in ref.~\cite{Kovtun:2012rj}, we can write the coefficients in terms of the elements of the susceptibility matrix in the grand canonical ensemble. The non-zero elements of the $3\times 3$ susceptibility matrix are $\chi_{11} = T (\partial \epsilon/\partial T)_{\mu/T}$, $\chi_{13} = \chi_{31} = (\partial \epsilon /\partial\mu)_{T}$, $\chi_{33} = (\partial n/\partial\mu)_{T}$, and $\chi_{22}=w_0$, with derivatives evaluated at constant $B^2$ in equilibrium. 
The longitudinal diffusion constant is
$$
  D_\parallel = \frac{\sigma_\parallel \, w_0^2}{n_0^2 \chi_{11} + w_0^2 \chi_{33} - 2 n_0 w_0 \chi_{13}}\,.
$$
The positivity of the diffusion constant implies $\sigma_\parallel >0$. The speed of sound squared expressed in terms of the elements of the susceptibility matrix is given by
$$
  v_s^2 = \frac{n_0^2 \chi_{11} + w_0^2 \chi_{33} -2n_0 w_0 \chi_{13}}{ \det(\chi)}\,,
$$
and the damping coefficient is
$$
  \Gamma_{s,\parallel} = \frac{1}{w_0} \left( \coeff43 (\eta_1 +\eta_2) + \zeta_1 +\zeta_2   \right) + 
  \frac{\sigma_\parallel\, w_0}{\det(\chi)} 
  \frac{(n_0 \chi_{11} - w_0 \chi_{13})^2}{n_0^2 \chi_{11} + w_0^2 \chi_{33} - 2n_0 w_0 \chi_{13}}
  \,.
$$
The expression for $v_s$ and $D_\parallel$ in terms of the thermodynamic functions formally look the same as in hydrodynamics without external $O(1)$ magnetic fields~\cite{Kovtun:2012rj}. All of $v_s$, $\Gamma_{s,\parallel}$, and $D_\parallel$ depend on $B_0$ through $p=p(T,\mu,B^2)$ and the transport coefficients.

For momenta ${\bf k}\perp {\bf B}_0$, the three gapless eigenmodes include two diffusive modes, and one ``subdiffusive'' mode with a quartic dispersion relation,
\begin{subequations}
\label{eq:omega-perp}
\begin{align}
  & \omega = -i D_\perp k^2\,,\\[5pt]
  & \omega = -i \frac{ \eta_\parallel k^2}{w_0}\,,\\[5pt]
  & \omega = -i \frac{\eta_\perp k^4}{B_0^2\, \chi_{33}}\,.
\end{align}
\end{subequations}
The transverse diffusion constant is determined by the transverse resistivity. We define the $2\times2$ conductivity matrix in the plane transverse to ${\bf B}_0$ as $\sigma_{ab} \equiv \sigma_\perp \delta_{ab} + \left(\frac{n_0}{|{\bf B}_0|} + \tilde\sigma \right)\epsilon_{ab}$, and the corresponding resistivity matrix as $\rho_{ab} \equiv (\sigma^{-1})_{ab} = \rho_\perp \delta_{ab} + \tilde\rho_\perp\, \epsilon_{ab}$, which defines $\rho_\perp$ and~$\tilde\rho_\perp$. The transverse diffusion constant is then
$$
  D_\perp = \frac{w_0^3 \chi_{33}}{\det(\chi) B_0^2} \, \rho_\perp\,,
$$
again using ${\MO}_{,\mu} B_0^2 \ll w_0$.
Stability of the equilibrium state now implies $\eta_\perp >0$, $\eta_\parallel >0$.

For modes propagating at an angle $\theta$ with respect to ${\bf B}_0$, the gapless modes include sound waves (unless $\theta=\pi/2$), and a diffusive mode. For a fixed value of $\theta$, the small-momentum eigenfrequencies are $\omega = \pm k v_s \cos\theta -\frac{i}{2}\Gamma_s(\theta) k^2$, and $\omega=-iD(\theta)k^2$, where
\begin{align*}
  & D(\theta) = D_\parallel \cos^2\theta + \frac{n_0^2}{v_s^2 w_0 \chi_{33}} D_\perp \sin^2\theta\,,\\[5pt]
  & \Gamma_{s}(\theta) = \Gamma_{s,\parallel} \cos^2\theta + \left( \frac{\eta_\parallel}{w_0} + \frac{(n_0 \chi_{13} - w_0 \chi_{33})^2}{\chi_{33}\, v_s^2\,\det(\chi)} D_\perp \right) \sin^2\theta\,.
\end{align*}
The coefficient $D_c$ in the cyclotron mode eigenfrequency (\ref{eq:omega-cyclo}) at small $B_0$ is 
$$
  D_c =  \left( \pm\frac{i v_s^2 w_0}{2 n_0 B_0} + \frac{(n_0^2 \chi_{11} {-} w_0^2 \chi_{33})w_0}{2 n_0^2 \det(\chi)}\sigma + \frac{3\zeta{+}7\eta}{6w_0} \right) \sin^2\theta + \frac{\eta}{w_0} \cos^2\theta + O(B_0)\,.
$$
Note that the limits $\theta\to \pi/2$ and $k\to0$ in the eigenfrequencies do not commute.

\subsection{Entropy production}
\label{sec:entropy-production}
The simple flat-space eigenfrequency analysis in the previous subsection imposes certain constraints on non-equilibrium transport coefficients. In order to find more general constraints, one method is to impose a local version of the second law of thermodynamics: the existence of a local entropy current with positive semi-definite divergence for every non-equilibrium configuration consistent with the hydrodynamic equations. We will not attempt to construct the most general entropy current from scratch. Rather, we will use the result of \cite{Bhattacharyya:2013lha, Bhattacharyya:2014bha} saying that the constraints on transport coefficients derived from the entropy current are the same as those derived from the equilibrium generating functional, plus the inequality constraints on dissipative transport coefficients. We take the entropy current to be
$$
  S^\mu = S^\mu_{\rm canon} + S^\mu_{\rm eq.}\,,
$$
where the canonical part of the entropy current is
\begin{equation}
\label{eq:S-canon}
S^\mu_{\rm canon} = \frac{1}{T} \left(p u^\mu - T^{\mu\nu}u_\nu - \mu J^\mu \right)\,,
\end{equation}
and $S^\mu_{\rm eq.}$ is found from the equilibrium partition function, as described in~\cite{Bhattacharyya:2013lha, Bhattacharyya:2014bha}. The constraints on transport coefficients follow by demanding $\nabla_{\!\mu} S^\mu \geqslant0$.
Using conservation laws (\ref{eq:TJ0}), the divergence of the canonical entropy current is
$$
\nabla_{\!\mu} S^\mu_{\rm canon} = \nabla_{\!\mu} \left(\frac{p}{T}u^\mu\right) - T^{\mu\nu}\nabla_{\!\mu} \frac{u_\nu}{T} + J^\mu\left(\frac{E_\mu}{T}-\partial_\mu \frac{\mu}{T}\right) \,.
$$
The $S^\mu_{\rm eq.}$ part of the entropy current is explicitly built to cancel out the part of $\nabla_{\!\mu} S^\mu_{\rm canon}$ that arises from the equilibrium terms in the constitutive relations, i.e.~the terms in $T^{\mu\nu}$ and~$J^\mu$ derived from the equilibrium generating functional. In fact, ref.~\cite{Bhattacharyya:2014bha} has already found $S^\mu_{\rm eq.}$ in the case when the generating functional contains a contribution proportional to $B{\cdot}\Omega$. We thus focus on non-equilibrium terms, and write the thermodynamic frame constitutive relations (\ref{eq:CRTF}) as $T^{\mu\nu} = T^{\mu\nu}_{\rm eq.} + T^{\mu\nu}_\textrm{non-eq.}$ and $J^{\mu} = J^{\mu}_{\rm eq.} + J^{\mu}_\textrm{non-eq.}$. The divergence of the entropy current is then
\begin{align*}
  \nabla_{\!\mu} S^\mu 
  & = \frac{1}{T}J^\mu_\textrm{non-eq.}\left( E_\mu-T\partial_\mu \frac{\mu}{T}\right) 
    - T^{\mu\nu}_\textrm{non-eq.} \nabla_{\!\mu} \frac{u_\nu}{T}\\[5pt]
  & = \frac{1}{T}\left( \ell^\mu_{\perp \textrm{non-eq.}} + \frac{B^\mu}{B}\ell_\textrm{non-eq.}\right) V_{\mu} -\frac{1}{T} f_\textrm{non-eq.} \nabla{\cdot} u 
    - \frac{1}{2T} t^{\mu\nu}_\textrm{non-eq.} \sigma_{\mu\nu} \,.
\end{align*}
Using the constitutive relations (\ref{eq:TTF}), (\ref{eq:JTF}), this leads to
\begin{align}
\label{eq:DS2}
  T \nabla_{\!\mu} S^\mu 
  & = \sigma_\parallel \frac{(B{\cdot} V)^2}{B^2} 
    + \sigma_\perp (\mathbb{B}^{\mu\nu} V_{\nu})^2 
    + \coeff12 \eta_\perp (\sigma^{\mu\nu}_\perp)^2  
    + \eta_\parallel \Sigma^2 \nonumber\\[5pt]
  & + (\zeta_1 - \coeff23 \eta_1) S_3^2
    + 2\eta_2 S_4^2 + (2\eta_1 + \zeta_2 -\coeff23 \eta_2) S_3 S_4\,,
\end{align}
where again $S_3\equiv \nabla{\cdot}u$ and $S_4 \equiv b^\mu b^\nu \nabla_{\!\mu}u_\nu$.
Demanding $\nabla_{\!\mu}S^\mu\geqslant0$ now gives
\begin{subequations}
\label{eq:entropy-constraints}
\begin{align}
\label{eq:entropy-constraints-a}
  \sigma_\parallel \geqslant0\,,\ \ \ \ 
  \sigma_\perp \geqslant0\,,\ \ \ \ 
  \eta_\perp \geqslant0\,,\ \ \ \ 
  \eta_\parallel \geqslant0\,,
\end{align}
together with the condition that the quadratic form made out of $S_3$, $S_4$ in the second line of eq.~(\ref{eq:DS2}) is non-negative, which implies
\begin{align}
\label{eq:entropy-constraints-b}
  & \eta_2 \geqslant 0\,,\ \ \ \ \ \ 
    \zeta_1 -\coeff23 \eta_1 \geqslant 0\,, \\[5pt]
\label{eq:S-constraint-2}
  & 2\eta_2 (\zeta_1 - \coeff23 \eta_1)
    \geqslant \coeff14
    (2\eta_1 + \zeta_2 -\coeff23 \eta_2)^2\,.
\end{align}
\end{subequations}
The coefficients $\tilde\eta_\perp$, $\tilde\eta_\parallel$, and $\tilde\sigma$ do not contribute to entropy production, and are not constrained by the above analysis. Thus, $\tilde\eta_\perp$, $\tilde\eta_\parallel$, and $\tilde\sigma$ are non-equilibrium non-dissipative coefficients.

\subsection{Kubo formulas}
\label{sec:correlators-1}
When the microscopic system is time-reversal invariant (i.e.~the only source of time-reversal breaking is due to the external magnetic field), transport coefficients can be further constrained by the Onsager relations. The retarded two-point functions of operators $O_a$ and $O_b$ in a time-reversal invariant theory in equilibrium obey
\begin{equation}
\label{eq:OR}
G_{ab}(\omega,\mathbf{k},B) = \epsilon_a \epsilon_b\, G_{ba}(\omega,-\mathbf{k},-B)\,,
\end{equation}
where $\epsilon_a$ and $\epsilon_b$ are time-reversal eigenvalues of the operators $O_a$ and $O_b$. We take our operators to be various components of $T^{\mu\nu}$ and $J^\mu$, and evaluate the retarded two-point functions by varying one-point functions in the presence of the external source with respect to the source. Namely, we solve the hydrodynamic equations in the presence of fluctuating external sources $\delta A, \delta g$ (proportional to $\exp(-i\omega t+ i{\bf k}{\cdot}{\bf x})$) to find $\delta T[A,g]$, $\delta \mu[A,g]$, $\delta u^\alpha[A,g]$, and then vary the resulting hydrodynamic expressions $T^{\mu\nu}[A,g]$ and $J^\mu[A,g]$ with respect to $g_{\alpha\beta}$, $A_\alpha$ to find the retarded functions. Specifically,
\begin{subequations}
\label{eq:corr-funcs}
\begin{align}
  & G_{T^{\mu\nu} T^{\alpha\beta}} = 2\frac{\delta}{\delta g_{\alpha\beta}} \left( \sqrt{-g}\, T^{\mu\nu}_\textrm{on-shell}[A,g] \right)\,,
  &&  G_{J^\mu T^{\alpha\beta}} = 2\frac{\delta}{\delta g_{\alpha\beta}} \left( \sqrt{-g}\, J^{\mu}_\textrm{on-shell}[A,g] \right)\,,\\[5pt]
  & G_{T^{\mu\nu} J^\alpha} = \frac{\delta}{\delta A_{\alpha}}   T^{\mu\nu}_\textrm{on-shell}[A,g]  \,,
  &&  G_{J^\mu J^\alpha} = \frac{\delta}{\delta A_{\alpha}}  J^{\mu}_\textrm{on-shell}[A,g] \,,
\end{align}
\end{subequations}
where the subscript ``on-shell'' signifies that the corresponding hydrodynamic $T^{\mu\nu}[A,g]$ and $J^\mu[A,g]$ are evaluated on the solutions to (\ref{eq:TJ0}), and the sources $\delta A$, $\delta g$ are set to zero after the variation is taken. The expressions (\ref{eq:corr-funcs}) are to be understood as 
$$
  \delta(\sqrt{-g}\, T^{\mu\nu}_\textrm{on-shell}) 
  = \coeff12 G_{T^{\mu\nu} T^{\alpha\beta}}(\omega, {\bf k})\, \delta g_{\alpha\beta} (\omega, {\bf k})\,,
$$
etc. This provides a direct method to evaluate the retarded functions, and allows both to check the Onsager relations and to derive Kubo formulas for transport coefficients.%
\footnote{
  Taken at face value, hydrodynamic correlation functions violate Onsager relations at non-zero $\omega$ and non-zero $k$. However these violations do not affect the Kubo formulas and disappear in the limit $B\ll T^2$, which corresponds to the validity regime of hydrodynamics.
}
The constraint on transport coefficients we find by demanding that eq.~(\ref{eq:OR}) holds is%
\footnote{
For parity-violating coefficients, we find $c_3 = \frac23 (c_{14} + c_{15}) - c_4$, $c_5 = -2(c_{14} + c_{15})$, $c_8 = -c_{15}$, $c_{10} = -c_{17}$.
}
\begin{equation}
\label{eq:OR2}
3 \zeta_2 - 6 \eta_1 - 2 \eta_2 = 0\,.
\end{equation}
For the rest of the paper, we will assume that (\ref{eq:OR2}) holds, which leaves us with ten non-equilibrium transport coefficients for a parity-invariant microscopic system.
Using eq.~(\ref{eq:OR2}) to eliminate $\zeta_2$, the inequality constraint in eq.~(\ref{eq:S-constraint-2}) turns into
\begin{equation}
   2\eta_2 (\zeta_1 - \coeff23 \eta_1)
    \geqslant 
    4\eta_1^2\,.
\end{equation}
We next list the expressions for transport coefficients in terms of retarded functions evaluated in flat-space equilibrium with external magnetic field in the $z$ direction, as in sec.~\ref{sec:eigenmodes-1}. In the limit ${\bf k}\to0$ first, $\omega\to 0$ second we find the following Kubo formulas. The two-point function of the longitudinal current $J^z$ gives the longitudinal conductivity,%
\begin{subequations}
\label{eq:Kubo-r1}
\begin{align}
\label{eq:Kubo-r1a}
   \coeff{1}{\omega} {\rm Im}\, G_{J^z J^z}(\omega,{\bf k}{=}0) =  \sigma_\parallel \,,
\end{align}
while the two-point functions of the transverse currents $J^x$, $J^y$ give the transverse resistivities,
\begin{align}
\label{eq:Kubo-r1b}
  &  \coeff{1}{\omega} {\rm Im}\, G_{J^x J^x}(\omega,{\bf k}{=}0) = \omega^2 \rho_\perp \frac{w_0^2}{B_0^4}  \,,\\
\label{eq:Kubo-r1c}
  &  \coeff{1}{\omega}{\rm Im}\, G_{J^x J^y}(\omega,{\bf k}{=}0) = 
     \frac{n_0}{B_0} - \omega^2 \tilde\rho_\perp \frac{w_0^2}{B_0^4}\, {\rm sign}(B_0)\,,
\end{align}
\end{subequations}
where the resistivities $\rho_\perp$ and $\tilde\rho_\perp$ were defined below eq.~(\ref{eq:omega-perp}).
Alternatively, the resistivities can be found from correlation functions of momentum density,%
\begin{subequations}
\label{eq:Kubo-r2}
\begin{align}
  & \coeff{1}{\omega} {\rm Im}\, G_{T_{0x} T_{0x}} (\omega, {\bf k}{=}0) = \rho_\perp \frac{w_0^2}{B_0^2} \,,\\[5pt]
  & \coeff{1}{\omega} {\rm Im}\, G_{T_{0x} T_{0y}} (\omega, {\bf k}{=}0) = -\tilde \rho_\perp {\rm sign}(B_0) \frac{w_0^2}{B_0^2}\,,
\end{align}
\end{subequations}
assuming $B_0^2\ll w_0$. The shear viscosities are given by 
\begin{subequations}
\label{eq:Kubo-Tmn}
\begin{align}
  & \coeff{1}{\omega}{\rm Im}\, G_{T^{xy} T^{xy}}(\omega,{\bf k}{=}0) =  \eta_\perp \,,\\
  & \coeff{1}{\omega}{\rm Im}\, G_{T^{xy} T^{xx}}(\omega,{\bf k}{=}0) =  \tilde\eta_\perp\, {\rm sign}(B_0) \,,\\
  & \coeff{1}{\omega}{\rm Im}\, G_{T^{xz} T^{xz}}(\omega,{\bf k}{=}0) =  \eta_\parallel \,,\\
  & \coeff{1}{\omega}{\rm Im}\, G_{T^{yz} T^{xz}}(\omega,{\bf k}{=}0) =  \tilde\eta_\parallel\, {\rm sign}(B_0)\,,
\end{align}
while the ``bulk'' viscosities may be expressed as
\begin{align}
  & \coeff{1}{\omega}\delta_{ij} {\rm Im}\, G_{T^{ij} T^{xx}}(\omega,{\bf k}{=}0) = 3 \zeta_1\,,\\
  & \coeff{1}{3\omega} \delta_{ij} \delta_{kl}\, {\rm Im}\, G_{T^{ij} T^{kl}}(\omega,{\bf k}{=}0) = 3\zeta_1 + \zeta_2 \,,\\
  & \coeff{1}{\omega} {\rm Im}\, G_{O_1 O_1} = \zeta_1 - \coeff23 \eta_1\,,\\
  & \coeff{1}{\omega} {\rm Im}\, G_{O_2 O_2} = 2\eta_2\,,
\end{align}
\end{subequations}
where $O_1 = \frac12(T^{xx}+T^{yy})$, and $O_2 = T^{zz} - \frac12(T^{xx}+T^{yy})$.
Correlation functions at non-zero momentum may be obtained in a straightforward way from the variational procedure described earlier.

\subsection{Inequality constraints on transport coefficients}
Finally, let us show that the inequality constraints on transport coefficients  derived from demanding that the entropy production is non-negative can also be obtained from hydrodynamic correlation functions, without using the entropy current. The argument is based on the fact that the imaginary part of the retarded function $G_{OO}(\omega,{\bf k})$ must be positive for any Hermitean operator~$O$ and $\omega>0$,
\begin{equation}
\label{eq:inequality-1}
  {\rm Im}\, G_{OO}(\omega, {\bf k}) \geqslant 0\,.
\end{equation}
Now consider the operator $O=a O_1 + b O_2$, with real coefficients $a$ and $b$, and Hermitean operators $O_{1}$, $O_2$. The inequality (\ref{eq:inequality-1}) implies 
$$
  {\rm Im} \left[ a^2 G_{O_1 O_1} + ab G_{O_1 O_2} + ab G_{O_2 O_1} + b^2 G_{O_2 O_2}\right] \geqslant 0\,,
$$
for $\omega\geqslant0$. This quadratic form in $a$, $b$ must be non-negative for all $a,b$ which implies ${\rm Im} G_{O_1 O_1} \geqslant 0$, ${\rm Im} G_{O_2 O_2} \geqslant 0$ together with
\begin{align}
\label{eq:inequality-2}
  \left( {\rm Im} G_{O_1 O_1}\right) \left({\rm Im} G_{O_2 O_2}\right) \geqslant \coeff14
  \left( {\rm Im} G_{O_1 O_2} + {\rm Im} G_{O_2 O_1} \right)^2\,.
\end{align}
The two terms in the right-hand side of (\ref{eq:inequality-2}) can be related by the Onsager relation (\ref{eq:OR}). As an example, take $O_1 = \frac12(T^{xx}+T^{yy})$, and $O_2 = T^{zz} - \frac12(T^{xx}+T^{yy})$. Evaluating the correlation functions at ${\bf k}=0$ and $\omega\to0$, the inequalities (\ref{eq:inequality-1}), (\ref{eq:inequality-2}) immediately imply the entropy current constraint~(\ref{eq:S-constraint-2}). The constraints (\ref{eq:entropy-constraints-a}), (\ref{eq:entropy-constraints-b}) follow directly from the Kubo formulas given in the previous subsection.

\section{Hydrodynamics with dynamical electromagnetic fields}
\label{sec:hydro-2}
\subsection{Dynamical gauge field}
We now move on to systems where the gauge field $A_\mu$ is dynamical rather than external, which will lead us to MHD. In external metric $g$, the (microscopic) generating functional is
$$
  Z[g] = \int\!\! DA \; e^{i S[g,A]}\,,
$$
where $S$ is the action. Let us couple the gauge field to an external conserved current $J^\mu_{\rm ext}$. We do this so that the new generating functional is
\begin{equation}
\label{eq:ZZ}
  Z[g,J_{\rm ext}] = \int\!\! DA\, D\varphi\; e^{iS[g,A] 
  + i\int\!\sqrt{-g}\, (A_\mu - \partial_\mu \varphi) J^\mu_{\rm ext}}\,,
\end{equation}
and $W\equiv-i\ln Z$.
The new field $\varphi$ is a Lagrange multiplier which shifts under gauge transformations and ensures that the external current is conserved. We define the energy-momentum tensor and the current by the variation of the action: 
$$
  \delta_g S[g,A] = \coeff12 \int\!\! \sqrt{-g}\, T^{\mu\nu} \delta g_{\mu\nu}\,,\ \ \ \ 
  \delta_A S[g,A] = \int\!\! \sqrt{-g}\, J^{\mu} \delta A_{\mu}\,.
$$
Diffeomorphism invariance of $W[g,J_{\rm ext}]$ implies
$
  \nabla_{\!\mu} \langle T^{\mu\nu}\rangle = 
  \langle F^{\lambda\nu} \rangle J_{{\rm ext}\,\lambda}\,.
$
In what follows, we will omit the angular brackets, writing the (non)-conservation of the energy-momentum tensor simply as 
\begin{equation}
\label{eq:T2}
  \nabla_{\!\mu}  T^{\mu\nu} = 
   F^{\lambda\nu} J_{{\rm ext}\,\lambda}\,.
\end{equation}
In the standard hydrodynamic approach, $T^{\mu\nu}$ and $F_{\mu\nu}$ will then be taken as dynamical variables in the classical hydrodynamic theory. 
Note that the sign in the right-hand side of eq.~(\ref{eq:T2}) is opposite compared to eq.~(\ref{eq:TJ1}), owing to the fact that the current, rather than the gauge field, is now external. 
In order to proceed with hydrodynamics, we need to specify {\it a)} the constitutive relations for the energy-momentum tensor to be used in eq.~(\ref{eq:T2}), and {\it b)} the equations which determine the evolution of the dynamical gauge field $F_{\mu\nu}$.

\subsection{Maxwell's equations in matter}
Classical equations specifying the dynamics of electric and magnetic fields are usually referred to as Maxwell's equations in matter. While we don't have a recipe of deriving them in a most general form in a model-independent way, a useful starting point is provided by matter in thermal equilibrium. Maxwell's equations for equilibrium matter may be then amended to include the non-equilibrium and dissipative effects, such as the electrical conductivity. To this end, as advocated in \cite{Jensen:2013vta}, we take the static generating functional $W_{\!s}[g,A]$ to be the effective action for gauge fields in equilibrium,
\begin{equation}
\label{eq:W0}
  S_{\rm eff}[g,A] = \int\!\!d^{4}x\;\sqrt{-g}\, {\cal F}\,,
\end{equation}
where ${\cal F}$ is a local gauge-invariant function of the sources $g_{\mu\nu}$ and $A_\mu$, and we have ignored the surface terms. To leading order in the derivative expansion, ${\cal F}$ is simply the pressure. We can always write ${\cal F}=-\coeff14 F_{\mu\nu}F^{\mu\nu} + {\cal F}_{\!\m}$, where the vacuum action is $-\coeff14 F_{\mu\nu} F^{\mu\nu} = \coeff12(E^2 - B^2)$, and ${\cal F}_{\!\m}$ is the ``matter" contribution. The isolation of the vacuum term is arbitrary, but it will allow us to make contact with the textbook form of Maxwell's equations in matter.
Our (equilibrium) effective theory is then given by the partition function (\ref{eq:ZZ}), with $S$ replaced by $S_{\rm eff}$, and the total action is
$$
  S_{\rm tot}[A, \varphi] = W_{\!s}[g,A] 
  + \int\!\!\sqrt{-g}\, (A_\mu {-} \partial_\mu \varphi) J^\mu_{\rm ext}\,.
$$
The current derived by varying the total action with respect to $A_\mu$ is ${J}^\mu_{\rm tot} = J^\mu + J^\mu_{\rm ext}$, or
$$
  {J}^\mu_{\rm tot} = -\nabla_{\!\nu} ( F^{\mu\nu} - M_\m^{\mu\nu}) + n u^\mu + J^\mu_{\rm ext} \,,
$$
where the polarization tensor $M_\m^{\mu\nu}$ is defined by 
$
  \delta_F \int\! d^4x\,\sqrt{-g}\, {\cal F}_\m = 
  \coeff12 \int\! d^4x\,\sqrt{-g}\, M_\m^{\mu\nu} \, \delta F_{\mu\nu}\,,
$
and the density of ``free'' charges is $n \equiv \partial {\cal F}_\m/\partial\mu$.
The equation of motion for the gauge field follows from $\delta_A S_{\rm tot}=0$, or equivalently ${J}^\mu_{\rm tot}=0$, and becomes
\begin{align}
\label{eq:MM0}
  \nabla_{\!\nu} H^{\mu\nu} = n u^\mu + J^\mu_{\rm ext}\,,
\end{align}
where $H^{\mu\nu}\equiv F^{\mu\nu} - M_\m^{\mu\nu}$. This is the desired equation that must be satisfied by electromagnetic fields in equilibrium. Following the standard hydrodynamic lore and assuming that eq.~(\ref{eq:MM0}) also holds for small departures away from equilibrium, one obtains hydrodynamics of ``perfect fluids'', now with dynamical electric and magnetic fields. 
For these perfect fluids, equations~(\ref{eq:MM0}) have to be solved together with the stress tensor (non)-conservation~(\ref{eq:T2}), where $T^{\mu\nu}$ is derived from the effective action~(\ref{eq:W0}).

In fact, eq.~(\ref{eq:MM0}) is nothing but the standard Maxwell's equations in matter.
The polarization tensor $M_\m^{\mu\nu}$ defines electric and magnetic polarization vectors $P^\mu$ and $M^\mu$ through the decomposition
\begin{equation}
\label{eq:M-tensor}
  M_\m^{\mu\nu} = P^\mu u^\nu - P^\nu u^\mu - \epsilon^{\mu\nu\rho\sigma} u_\rho M_{\sigma}\,.
\end{equation}
The antisymmetric tensor $H_{\mu\nu}$ can be decomposed in the same way as the field strength $F_{\mu\nu}$,
$$
  H_{\mu\nu} = u_\mu D_\nu - u_\nu D_\mu - \epsilon_{\mu\nu\rho\sigma} u^\rho H^\sigma\,,
$$
which defines $D_\mu\equiv H_{\mu\nu}u^\nu$ and $H^\mu \equiv \coeff12 \epsilon^{\mu\nu\alpha\beta} u_\nu H_{\alpha\beta}$, so that
\begin{align*}
  & D^\mu = E^\mu + P^\mu\,,\\[5pt]
  & H^\mu = B^\mu - M^\mu\,.
\end{align*}
It is then clear that eq.~(\ref{eq:MM0}) is the covariant form of Maxwell's equations in matter: the currents of `free charges' are in the right-hand side, while the effects of polarization appear in the left-hand side through the substitution $E^\mu\to D^\mu$, $B^\mu\to H^\mu$ in the vacuum Maxwell's equations. Action~(\ref{eq:W0}) is the action for Maxwell's equations in matter.

As an example, consider the following ``matter'' contribution: ${\cal F}_{\m} = p_{\!\m}(T,\mu,E^2,B^2,E{\cdot}B)$, where $p_{\!\m}$ is the ``matter'' pressure. The polarization tensor is then $M^{\mu\nu}_\m = 2\partial p_{\!\m}/\partial F_{\mu\nu}$, and the polarization vectors are
\begin{subequations}
\label{eq:PM-vectors}
\begin{align}
  & P^\mu = \cEE E^\mu + \cEB B^\mu  \,,\\[5pt]
  & M^\mu = \cEB E^\mu + \cBB B^\mu  \,,
\end{align}
\end{subequations}
where the susceptibilities $\cEE \equiv 2\partial p_\m/\partial E^2$, $\cEB \equiv \partial p_\m/\partial(E{\cdot}B)$, and $\cBB \equiv 2\partial p_\m/\partial B^2$ all depend on $T$, $\mu$, $E^2$, $B^2$, and $E{\cdot}B$. This gives the standard constitutive relations, expressing $D$ and $B$ in terms of $E$ and $H$,
\begin{align*}
  & D^\mu = \varepsilon_\m E^\mu + \beta_\m H^\mu \,,\\[5pt]
  & B^\mu = \beta_\m E^\mu + \mu_\m H^\mu  \,,
\end{align*}
where $\varepsilon_\m \equiv 1+\cEE + \cEB^2/(1{-}\cBB)$ is the electric permittivity, $\mu_\m \equiv 1/(1{-}\cBB)$ is the magnetic permeability, and $\beta_\m \equiv \cEB/(1{-}\cBB)$.
We will also use $\varepsilon_\e \equiv 1{+}\cEE$, which coincides with the electric permittivity if $\cEB=0$.

\subsection{Hydrodynamics}
\label{sec:MHD-eqs}
We take the MHD equations to be as follows:
\begin{subequations}
\label{eq:hydro-eqs}
\begin{align}
\label{eq:TC2}
  & \nabla_{\!\mu}  T^{\mu\nu} = F^{\lambda\nu} J_{{\rm ext}\,\lambda}\,, \\[5pt]
\label{eq:ME2}
  & J^\mu + J^\mu_{\rm ext} = 0\,,\\[5pt]
\label{eq:BI}
  & \epsilon^{\mu\nu\alpha\beta} \nabla_{\!\nu} F_{\alpha\beta} = 0\,.
\end{align}
\end{subequations}
The last equation is the electromagnetic ``Bianchi identity'', expressing the fact that the electric and magnetic fields are derived from the vector potential $A_\mu$. The second equation (Maxwell's equations in matter) can be rewritten as $\nabla_{\!\nu} (F^{\mu\nu} {-} M_\m^{\mu\nu}) = J^\mu_{\rm free} + J^\mu_{\rm ext}$ which defines $J^\mu_{\rm free}$, the current of ``free charges''. While eqs.~(\ref{eq:TC2}) and (\ref{eq:BI}) are true microscopically, the Maxwell's equations in matter (\ref{eq:ME2}) are written based on the above intuition of the equilibrium effective action. Note that $\nabla_{\!\mu} J^\mu_{\rm free} =0$ is a consequence of (\ref{eq:ME2}), and is not an independent equation. The hydrodynamic variables are $T$, $u^\alpha$, $\mu$, as well as the electric and magnetic fields which satisfy $u_\alpha E^\alpha=0$, $u_\alpha B^\alpha=0$. Hydrodynamic equations (\ref{eq:hydro-eqs}) must be supplemented by constitutive relations, which express $T^{\mu\nu}$, $J^\mu$ (or $J^\mu_{\rm free}$ and $M^{\mu\nu}_{\!\m}$) in terms of the hydrodynamic variables. These constitutive relations will contain equilibrium contributions coming from the equilibrium effective action~(\ref{eq:W0}). In addition, the constitutive relations will contain non-equilibrium contributions, such as the electrical conductivity and the shear viscosity. 

Taking the divergence of eq.~(\ref{eq:ME2}) and using $J^\mu_{\rm ext} = -J^\mu$ gives
\begin{align*}
  & \nabla_{\!\mu}  T^{\mu\nu} = F^{\nu\lambda} J_{\lambda}\,, \\
  & \nabla_{\!\mu} J^\mu = 0\,,
\end{align*}
which shows that the variables $T$, $u^\alpha$, and $\mu$ satisfy exactly the same equations (\ref{eq:TJ0}) as they did in the theory with a non-dynamical, external $A_\mu$. Thus in order to ``solve'' the MHD theory (\ref{eq:hydro-eqs}) one can {\it i)} solve the hydrodynamic equations with an external gauge field (\ref{eq:hydro-eqs}) to find $T[A,g]$, $u^\alpha[A,g]$, $\mu[A,g]$, and {\it ii)} solve $J^\mu [T[A,g], u^\alpha[A,g], \mu[A,g],A,g] + J^\mu_{\rm ext} = 0$ in order to find $A_\mu[J_{\rm ext},g]$, and {\it iii)} use the constitutive relations to find the energy-momentum tensor $T^{\mu\nu}[J_{\rm ext},g] = T^{\mu\nu}[T[A[J_{\rm ext},g],g], u^\alpha[A[J_{\rm ext},g],g], \mu[A[J_{\rm ext},g],g], A[J_{\rm ext},g], g ]$. MHD correlation functions may then be obtained through variations with respect to the external sources $J^\lambda_{\rm ext}$ and $g_{\mu\nu}$.

An equivalent way to understand the classical effective theory (\ref{eq:hydro-eqs}) is to promote the real-time generating functional to the non-equilibrium effective action~\cite{Jensen:2013vta}, i.e.\ to write
$$
  S_{\rm tot}[A, \varphi] = W_{\!r}[A,g] 
  + \int\!\!\sqrt{-g}\, (A_\mu {-} \partial_\mu \varphi) J^\mu_{\rm ext}\,,
$$
where $W_r[A,g]$ is low-energy, real-time generating functional for retarded correlation functions in the theory with a non-dynamical $A_\mu$. The functional $W_r[g,A]$ is non-local due to the gapless low-energy degrees of freedom (sound waves etc). However, for the purposes of MHD we do not need the actual generating functional, but only the equations of motion for the effective action $S_{\rm tot}$. These equations of motion are $J^\mu[A,g] + J^\mu_{\rm ext} = 0$, where $J^\mu[A,g]$ is the on-shell current in the theory with a non-dynamical $A_\mu$. One can then solve the theory as described in the previous paragraph.

We will thus adopt the simplest hydrodynamic effective theory~(\ref{eq:hydro-eqs}) where the constitutive relations for $T^{\mu\nu}$ and $J^\mu$ are the same as in the case of external non-dynamical electromagnetic fields. Under this ``mean-field'' assumption, transport coefficients which are naively independent would still be related by the conditions originating from the static generating functional. 

Further, any solution $T[A,g]$, $u^\alpha[A,g]$, $\mu[A,g]$ to the MHD equations is also a solution to the hydrodynamic equations (\ref{eq:TJ0}) in the theory with a non-dynamical $A_\mu$. Thus the entropy current with a non-negative divergence on the solutions to (\ref{eq:TJ0}) will also have non-negative divergence when evaluated on the solutions to the MHD equations~(\ref{eq:hydro-eqs}). This means that the entropy current in MHD may be taken the same as the entropy current in the theory with a non-dynamical gauge field~\cite{Jensen:2013vta}, and we do not need to perform a separate entropy current analysis beyond what was already done in sec.~\ref{sec:hydro-1}.

To sum up, with the MHD scaling $B\sim O(1)$, $E\sim O(\partial)$, the equilibrium effective action is given by eq.~(\ref{eq:F1}),
\begin{equation}
\label{eq:F2}
 S_{\rm eff} = \int\!\! \sqrt{-g} \left( -\coeff12 B^2 + p_{\!\m}(T,\mu, B^2) + \sum_{n=1}^{5} M_n(T,\mu,B^2) s_n^{(1)} + O(\partial^2) \right)\,.
\end{equation}
For a parity-invariant theory, only the $M_4$ term in the sum contributes.
The constitutive relations for the energy-momentum tensor and the current were already found in the previous section, where now we have $p(T,\mu,B^2) = -\coeff12 B^2 + p_{\!\m}(T,\mu, B^2)$.
The energy-momentum tensor appearing in eq.~(\ref{eq:hydro-eqs}) and the current $J^\mu$ satisfying $J^\mu + J^\mu_{\rm ext} = 0$ take the form (\ref{eq:TT1}), (\ref{eq:JJ1}), and the constitutive relations for a parity-invariant theory in the thermodynamic frame are given by Eqs.~(\ref{eq:TTF}), (\ref{eq:JTF}).

We will find it useful to modify the above effective theory by giving dynamics to the electric field. To do so, we add an $O(\partial^2)$ term $\frac12 \varepsilon_\e E^2$ to the effective action~(\ref{eq:F2}), where $\varepsilon_\e$ is the electric permittivity which we take constant. This term is one of the many $O(\partial^2)$ terms, and we add it as a ``ultraviolet regulator'' which improves the high-frequency behaviour of the theory. When studying the near-equilibrium eigenmodes of the system, this term will affect the frequency gaps, but not the leading-order dispersion relations of the gapless modes. With this new term, the following contributions have to be added to the constitutive relations (\ref{eq:TTF}), (\ref{eq:JTF}):
\begin{align*}
  & T^{\mu\nu}_{\rm El.} = \varepsilon_\e \left(\coeff12 E^2 g^{\mu\nu} + E^2 u^\mu u^\nu - E^\mu E^\nu \right)\,,\\
  & J^\mu_{\rm El.} = -\varepsilon_\e \nabla_{\lambda}\left( E^\lambda u^\mu - E^\mu u^\lambda \right)\,.
\end{align*}
The current $J^\mu_{\rm El.}$ contains the kinetic term for the electric field in Maxwell's equations, as well as the ``bound'' current due to electric polarization.

\subsection{Eigenmodes}
\label{sec:eigenmodes-2}
As a simple application of the above MHD theory, one can study the eigenmodes of small oscillations about the thermal equilibrium state. As we did earlier, we set the external sources to zero, and linearize the hydrodynamic equations near the flat-space equilibrium state with constant $T=T_0$, $\mu=\mu_0$, $u^\alpha=(1,{\bf 0})$, and $B^\alpha=(0,0,0,B_0)$. For simplicity, we will take the magnetic permeability~$\mu_\m$ constant, though it is straightforward to find how the eigenfrequencies below are modified for non-constant $\mu_\m = \mu_\m(T,\mu,B^2)$.

\subsubsection*{Neutral state}
We begin with the neutral state at $\mu_0 =0$ and $n_0 = 0$. The system admits nine eigenmodes, three gapped, and six gapless.

Let us start with the familiar case of vanishing magnetic field in equilibrium. The system is then isotropic, with shear viscosity $\eta$, bulk viscosity $\zeta$, and conductivity $\sigma\equiv \sigma_\perp = \sigma_\parallel$. The fluctuations of $\delta T$, $\delta u_i$ decouple from the fluctuations of $\delta\mu$, $\delta E_i$, $\delta B_i$. The eigenmodes include two transverse shear modes with eigenfrequency $\omega=-i\eta k^2/(\epsilon_0{+}p_0)$, and longitudinal sound waves with $v_s^2=\partial p/\partial\epsilon$ and $\Gamma_s=(\coeff43\eta+\zeta)/(\epsilon_0+p_0)$. In addition, there is a longitudinal charge diffusion mode which becomes gapped because of non-zero electrical conductivity,
$$
  \omega = -\frac{i\sigma}{\varepsilon_\e} - i\left(\frac{\sigma}{\partial n/\partial\mu}\right) k^2\,.
$$
Thus, charge fluctuations in a neutral conducting medium do not diffuse. Instead, what diffuses are the transverse magnetic and electric fields: there are two sets of transverse conductor modes whose eigenfrequencies are determined by
$$
  \omega \left( \omega + \frac{i\sigma}{\varepsilon_\e} \right) = \frac{k^2}{\varepsilon_\e \mu_\m}\,.
$$
Recall that $\varepsilon_\e$ is the electric permittivity and $\mu_\m = 1/(1{-}2\partial p_\m/\partial B^2)$ is the magnetic permeability, so $\sqrt{\varepsilon_\e \mu_\m}$ is the elementary index of refraction. The conductor modes have the following frequencies at small momenta:
\begin{align*}
   \omega = -\frac{i\sigma}{\varepsilon_\e} + \frac{ik^2}{\sigma \mu_\m}\,,
\ \ \ \ \ \ 
   \omega = -\frac{ik^2}{\sigma \mu_\m}\,.
\end{align*}
The gapless conductor mode is responsible for the skin effect in metals.

We now turn on non-zero magnetic field and consider modes propagating at an angle~$\theta$ with respect to ${\bf B}_0$. Thermal and mechanical fluctuations now no longer decouple from electromagnetic fluctuations.  There is one longitudinal gapped mode, and two transverse gapped modes,
\begin{align*}
   \omega = -\frac{i\sigma_\parallel}{\varepsilon_\e} 
    + O(k^2)\,,\ \ \ \ \ \ 
   \omega = -\frac{i\sigma_\perp \pm \tilde\sigma}{\varepsilon_\e} +O(k^2)\,.
\end{align*}
In writing down the transverse eigenfrequencies, we have assumed $B_0^2\ll \epsilon_0+p_0$. 

All six gapless modes have linear dispersion relation at small momenta. Two of the gapless modes are the Alfv\'en waves,
\begin{subequations}
\label{eq:Alfven-waves}
\begin{align}
  \omega = \pm v_{\rm A} k \cos\theta - \frac{i\Gamma_{\rm A}}{2} k^2\,,
\end{align}
whose speed and damping are determined by
\begin{align}
\label{eq:Alfven-VG}
  v_{\rm A}^2 = \frac{B_0^2}{ \mu_\m (\epsilon_0 {+} p_0) + B_0^2}\,,\ \ \ \ 
  \Gamma_{\rm A} = \frac{1}{\epsilon_0 {+}p_0} \left( \eta_\perp \sin^2\theta +\eta_\parallel \cos^2\theta \right)
  +\frac{1}{\mu_\m} \left( \frac{}{}\rho_\perp \cos^2\theta + \rho_\parallel \sin^2\theta \right) \,,
\end{align}
\end{subequations}
where $\rho_\parallel \equiv 1/\sigma_\parallel$, and $\rho_\perp$ was defined below eq.~(\ref{eq:omega-perp}). In writing down the damping coefficient, we have taken $B_0^2\ll \epsilon_0 {+} p_0$, the corrections of order $B_0^2/(\epsilon_0 {+} p_0)$ are straightforward to write down. The other four gapless modes are the two branches of magnetosonic waves,
\begin{subequations}
\label{eq:ms-waves}
\begin{align}
  \omega = \pm v_{\rm ms} k - \frac{i\Gamma_{\rm ms}}{2} k^2\,,
\end{align}
whose speed is determined by the quadratic equation
\begin{equation}
\label{eq:ms-speed}
  (v_{\rm ms}^2)^2 - v_{\rm ms}^2 (v_A^2 + v_s^2 - v_A^2 v_s^2 \sin^2\theta) + v_A^2 v_s^2 \cos^2\theta = 0\,,
\end{equation}
where $v_s^2 = (s/T)/(\partial s/\partial T) = \partial p/\partial\epsilon$ is the speed of sound at $n_0=0$. The two solutions of (\ref{eq:ms-speed}) correspond to the sound-type (or ``fast'') branch, and the Alfv\'en-type (or ``slow'') branch. At $\theta=0$, the slow branch turns into a second set of Alfv\'en waves, while the fast branch becomes the sound wave.
See e.g. ref.~\cite{PhysRev.108.1357} for an early derivation of $v_{\rm A}$ and $v_{\rm ms}$ in relativistic MHD.
The damping coefficients of the magnetosonic waves are straightforward to evaluate, but are quite lengthy to write down in general, and we will only present them in the limits of small $B_0$ and small~$\theta$. As $B_0\to0$, the damping coefficients become
\begin{align}
  \textrm{slow:     } & \Gamma_{\rm ms} = \frac{\eta}{\epsilon_0{+}p_0}  
  + \frac{1}{\sigma \mu_\m}\,,\\[5pt]
  \textrm{fast:     } & \Gamma_{\rm ms} = \frac{1}{\epsilon_0{+}p_0} 
  \left( \coeff43 \eta + \zeta \right)\,.
\end{align}
On the other hand, as $\theta\to0$, the damping coefficients become
\begin{align}
\label{eq:Gms-slow-theta0}
  \textrm{slow:     } & \Gamma_{\rm ms} = \frac{\eta_\parallel}{\epsilon_0{+}p_0}  
  + \frac{\rho_\perp}{\mu_\m}\,,\\[5pt]
\label{eq:Gms-fast-theta0}
  \textrm{fast:     } & \Gamma_{\rm ms} = \frac{1}{\epsilon_0{+}p_0} 
  \left( \coeff{10}{3}\eta_1 + 2\eta_2 + \zeta_1 \right)\,.
\end{align}
\end{subequations}
We have again taken $B_0^2\ll \epsilon_0 +p_0$, the corrections of order $B_0^2/(\epsilon_0 {+} p_0)$ are straightforward to write down.
At $\theta=0$, both polarizations of Alfv\'en waves have the same damping.

Let us now consider gapless modes propagating perpendicularly to the magnetic field, i.e.\ taking $\theta\to\pi/2$ first, $k\to0$ second. These include sound waves
\begin{subequations}
\begin{align}
  \omega = \pm k v_{\pi/2} - \frac{i\Gamma_{\pi/2}}{2}k^2\,,
\end{align}
where $v_{\pi/2}$ is the non-zero solution of eq.~(\ref{eq:ms-speed}) at $\theta=\pi/2$. In the limit of small $B_0$ it reduces to $v_{\pi/2}^2 = v_s^2 = (s/T)/(\partial s/\partial T)=\partial p/\partial\epsilon$, in equilibrium. The damping coefficient is 
\begin{align}
  \Gamma_{\pi/2} = \frac{1}{\epsilon_0 {+} p_0} \left( \zeta_1 - \coeff23 \eta_1 + \eta_\perp\right)\,,
\end{align}
\end{subequations}
assuming $B_0^2\ll \epsilon_0 {+} p_0$. The other four gapless modes at $\theta=\pi/2$ are purely diffusive,
\begin{subequations}
\begin{align}
  & \omega = -\frac{i \eta_\parallel}{\epsilon_0{+}p_0}k^2\,,\\[5pt]
  & \omega = -\frac{i \rho_\parallel}{\mu_\m}k^2\,,\\[5pt]
\label{eq:w-eta2}
  & \omega = -\frac{i\eta_\perp}{\epsilon_0{+}p_0}k^2\,,\\[5pt]
\label{eq:w-perp2}
  & \omega = -\frac{i\rho_\perp}{\mu_\m}k^2\,,
\end{align}
\end{subequations}
In writing down (\ref{eq:w-eta2}) and (\ref{eq:w-perp2}) we have again taken $B_0^2\ll \epsilon_0{+}p_0$.

\subsubsection*{Charged state offset by background charge}
We now consider a state with a non-zero value of $\mu_0$, which gives rise to a constant non-zero charge density $n_0$. In order to ensure that the equilibrium state is stable, we will offset this equilibrium value of the dynamical charge density by a constant non-dynamical external background charge density $-n_0$. This can be achieved by choosing the external current in the hydrodynamic equations (\ref{eq:hydro-eqs}) as $J^\mu_{\rm ext}=(-n_0,{\bf 0})$. In the particle language, this would correspond to a state where the excess of electrically charged particles over antiparticles (or vice versa) is compensated by a constant charge density of immobile background ``ions''.  Even though the system is overall electrically neutral, its dynamics is not equivalent to that of the system with $\mu_0=0$, $n_0=0$: for example, the fluctuation of the spatial electric current has a convective contribution $n_0\, \delta u_i$. More formally, when analyzing hydrodynamic modes, the limits $n_0\to0$ and $k\to0$ do not commute. We now find six gapped modes and three gapless modes.

To get some intuition about the gapped modes, let us set all transport coefficients to zero, as well as set $B_0=0$.
Then at small momenta there are two longitudinal gapped modes whose frequencies are determined by
$$
  \omega^2 = \Omega_p^2 + v_s^2 k^2\,,
$$
where $\Omega_p^2 \equiv {n_0^2}/{[(\epsilon_0{+}p_0)\varepsilon_\e]}$, and $v_s$ is the speed of sound that the charged fluid would have, if the electromagnetic fields were not dynamical, see Sec~\ref{sec:eigenmodes-1}.
These modes are the relativistic analogues of Langmuir oscillations, and $\Omega_p$ is the relativistic ``plasma frequency" which gaps out the sound waves. In addition, there are four transverse gapped modes whose frequencies are determined by
$$
  \omega^2 = \Omega_p^2 + \frac{k^2}{\varepsilon_\e \mu_\m} \,.
$$
These are electromagnetic waves in the fluid, gapped by the same plasma frequency $\Omega_p$ as the sound waves. If we now turn on the transport coefficients, the gaps are determined by 
\begin{align*}
   \omega \left( \omega + \frac{i\sigma_\parallel}{\varepsilon_\e}\right) = \Omega_p^2\,,
   \ \ \ \ \ \
   \omega \left( \omega + \frac{i(\sigma_\perp \pm i\tilde\sigma)}{\varepsilon_\e}\right) = \Omega_p^2\,,
\end{align*}
indicating the damping of plasma oscillations. At non-zero $B_0^2\ll \epsilon_0+p_0$, the gaps will receive dependence on the magnetic field. 

At $B_0=0$ the system is isotropic. The gapless modes ($B_0\to0$ first, $k\to0$ second) include two transverse shear modes with quartic dispersion relation, and one longitudinal diffusive mode,
\begin{align*}
   \omega = -\frac{i\eta k^4}{n_0^2 \mu_\m}\,,\ \ \ \ \ \ 
   \omega = -\frac{i\sigma \chi_{33} w_0^3}{n_0^2 \det(\chi)}k^2\,,
\end{align*}
where again $w_0\equiv T_0 s_0 + \mu_0 n_0$, and the susceptibility matrix $\chi$ was defined below eq.~(\ref{eq:omega-par}).

At non-zero $B_0$, the three gapless modes all have quadratic dispersion relation at small momenta. 
There are two propagating waves with real frequencies
\begin{align}
\label{eq:w2}
  & \omega = \pm \frac{B_0 \cos\theta}{n_0\mu_\m} \, k^2\,,
\end{align}
where $\theta$ is the angle between ${\bf k}$ and ${\bf B}_0$, and one diffusive mode. For $B_0^2 {\MO}_{,\mu}\ll \epsilon_0 + p_0$, the diffusive frequency is
\begin{align}
  & \omega = -i\frac{\chi_{33} w_0^3}{{\rm det}(\chi)} 
    \left( \frac{\sigma_\parallel \cos^2\theta}{n_0^2} + \frac{\rho_\perp \sin^2\theta}{B_0^2}\right) k^2\,.
\end{align}
For gapless modes propagating at $\theta=\pi/2$ at small momenta ($\theta\to\pi/2$ first, $k\to0$ second), we again find the diffusive mode $\omega=-iD_\perp k^2$, with the same coefficient $D_\perp$ as in sec.~\ref{sec:eigenmodes-1}. In addition, at $\theta=\pi/2$ there are two ``subdiffusive'' modes with quartic dispersion relation,
\begin{align*}
  \omega = -i\frac{\eta_\perp k^4}{n_0^2 \mu_\m}\,,\ \ \ \ \ \ 
  \omega = -i\frac{\eta_\parallel k^4}{n_0^2 \mu_\m}\,.
\end{align*}
The eigenfrequencies are noticeably different from the ones in a theory with fixed, non-dynamical electromagnetic field discussed in sec.~\ref{sec:eigenmodes-1}. Compared to the case of $n_0=0$ earlier in this section, one can say that non-vanishing dynamical charge density gaps out the magnetosonic waves, and turns Alfv\'en waves into waves whose frequency is quadratic in momentum.

\subsection{Kubo formulas}
\label{sec:correlators-2}
We can find MHD correlation functions following the same variational procedure outlined in sec.~\ref{sec:correlators-1}. As the total current vanishes by the equations of motion, the objects whose correlation functions it makes sense to evaluate in MHD are the energy-momentum tensor $T^{\mu\nu}$ and the electromagnetic field strength tensor $F_{\mu\nu}$. It is straightforward to evaluate retarded functions in flat space, in an equilibrium state with constant $T=T_0$, $\mu=\mu_0$, $u^\alpha=(1,{\bf 0})$, and constant magnetic field. 
We solve the hydrodynamic equations in the presence of fluctuating external sources $\delta J_{\rm ext}, \delta g$ (proportional to $\exp(-i\omega t+ i{\bf k}{\cdot}{\bf x})$) to find $\delta T[J_{\rm ext},g]$, $\delta \mu[J_{\rm ext},g]$, $\delta u^\alpha[J_{\rm ext},g]$, $\delta F_{\mu\nu}[J_{\rm ext},g]$ and then vary the resulting hydrodynamic expressions $T^{\mu\nu}[J_{\rm ext},g]$ and $F_{\mu\nu}[J_{\rm ext},g]$ with respect to $g_{\alpha\beta}$, $J_{\rm ext}^\alpha$ to find the retarded functions. The metric variations are performed as usual,
\begin{align*}
  & G_{T^{\mu\nu} T^{\alpha\beta}} = 2\frac{\delta}{\delta g_{\alpha\beta}} \left( \sqrt{-g}\, T^{\mu\nu}_\textrm{on-shell}[J_{\rm ext},g] \right)\,,
  &  G_{F_{\mu\nu} T^{\alpha\beta}} = 2\frac{\delta}{\delta g_{\alpha\beta}} \left( \sqrt{-g}\, F_{\mu\nu}^\textrm{on-shell}[J_{\rm ext},g] \right)\,.
\end{align*}
The subscript ``on-shell'' signifies that $T^{\mu\nu}$ and $F_{\mu\nu}$ are evaluated on the solutions to (\ref{eq:hydro-eqs}) with the constitutive relations~(\ref{eq:TTF}), (\ref{eq:JTF}).
Further, recall that the external current must be conserved, which can be implemented by choosing $\delta J^0_{\rm ext} = k_i\, \delta J^i_{\rm ext}/\omega + \frac12 n_0 \delta g_\mu^{\ \;\mu}$. The coupling $A_\mu J^\mu_{\rm ext}$ then implies that $i\omega\, \delta/\delta J^l_{\rm ext}(k)$ produces an insertion of $F_{0l}(-k)$, while $ik_m \epsilon^{nml} \delta/\delta J^l_{\rm ext}(k)$ produces an insertion of $\frac12 \epsilon^{nml}F_{lm}(-k)$. For example, for electric field correlation functions we have 
\begin{align*}
  & G_{T^{\mu\nu} F_{0l}} = i\omega \frac{\delta}{\delta J^l_{\rm ext}}  T^{\mu\nu}_\textrm{on-shell}[J_{\rm ext},g] \,,
  &  G_{F_{\mu\nu} F_{0l}} = i\omega \frac{\delta}{\delta J^l_{\rm ext}}  F_{\mu\nu}^\textrm{on-shell}[J_{\rm ext},g] \,,
\end{align*}
and similarly for the magnetic field.%
\footnote{%
Alternatively, one can introduce an antisymmetric ``polarization source'' ${M}_{\rm ext}^{\mu\nu}$, by taking the conserved current as $J^\mu_{\rm ext} = \nabla_{\!\nu} {M}_{\rm ext}^{\mu\nu}$. The coupling $A_\mu J^\mu_{\rm ext}$ then becomes $\frac12 {M}_{\rm ext}^{\mu\nu} F_{\mu\nu}$ upon integration by parts, and correlation functions of $F_{\mu\nu}$ may be obtained as variations with respect to ${M}_{\rm ext}^{\mu\nu}$.
}

Choosing the external magnetic field in the $z$-direction, we find the same Kubo formulas~(\ref{eq:Kubo-r2}) and (\ref{eq:Kubo-Tmn}). The electrical resistivities may also be expressed in terms of correlation functions of the electric field. In the zero-density state with $\mu_0=0$, $n_0=0$ we find%
\begin{subequations}
\label{eq:Kubo-E}
\begin{align}
  & \coeff{1}{\omega} {\rm Im}\, G_{F_{z0} F_{z0}} (\omega, {\bf k}{=}0) = \rho_\parallel\,,
\end{align}
at small frequency, where $\rho_\parallel \equiv 1/\sigma_\parallel$. Similarly, for the transverse resistivities we find
\begin{align}
\label{eq:Kubo-E-2}
  & \coeff{1}{\omega} {\rm Im}\, G_{F_{x0} F_{x0}} (\omega, {\bf k}{=}0) =  \rho_\perp
    \,,\\
\label{eq:Kubo-E-3}
  & \coeff{1}{\omega} {\rm Im}\, G_{F_{x0} F_{y0}} (\omega, {\bf k}{=}0) = - \tilde\rho_\perp\,
    {\rm sign}(B_0)
    \,,
\end{align}
\end{subequations}
where again $w_0\equiv \epsilon_0{+}p_0$, and  $\rho_\perp$, $\tilde\rho_\perp$ were defined below eq.~(\ref{eq:omega-perp}). We have taken $B_0^2\ll w_0$, otherwise there is a multiplicative factor of ${w_0 (w_0 {-} B_0^2 {\MO}_{,\mu}) \mu_\m^2}/{(w_0 \mu_\m {+} B_0^2)^2}$ in the right-hand side of (\ref{eq:Kubo-E-2}), (\ref{eq:Kubo-E-3}). %
In a charged state (offset by non-dynamical $-n_0$), the correlation functions change, for example
$
  G_{F_{x0} F_{y0}}(\omega, {\bf k}{=}0) = i\omega \frac{B_0}{n_0}\,,
$
while $\sigma_\parallel$ can be found from%
\begin{align}
  \coeff{1}{\omega} {\rm Im}\, G_{T_{0z} T_{0z}} (\omega, {\bf k}{=}0) = \sigma_\parallel\,.\end{align}
Retarded functions at non-zero momentum may be found from the above variational procedure. For example, the function $G_{F_{x0} F_{x0}} (\omega, {\bf k})$ in a state with $n_0=0$ and with ${\bf k} \parallel {\bf B}_0$ has singularities at the eigenfrequencies of Alfv\'en waves for small momenta.

\section{A dual formulation}
\label{sec:comparison-GHI}
As this paper was being completed, an interesting article~\cite{Grozdanov:2016tdf} (abbreviated below as GHI) came out which approached magnetohydrodynamics from a different perspective. The dual electromagnetic field strength tensor $J^{\mu\nu} \equiv \frac12 \epsilon^{\mu\nu\alpha\beta}F_{\alpha\beta}$ was taken as a conserved current, and the constitutive relations were written down for $J^{\mu\nu}$, rather than for the electric current $J^\mu$ as was done in MHD historically. This ``dual'' construction follows the earlier work of ref.~\cite{Schubring:2014iwa} which studied a similar MHD-like setup for ``string fluids''. The paper \cite{Grozdanov:2016tdf} identifies six transport coefficients in MHD, compared to eleven transport coefficients (in a parity-preserving system) found here. In this section we revisit the analysis of GHI, and show that the dual formulation allows for the same eleven transport coefficients we described earlier in Sections~\ref{sec:hydro-1} and \ref{sec:hydro-2}.

\subsection{Constitutive relations}
The conservation laws are taken as follows:
\begin{equation}
\label{eq:GHI-eom}
  \nabla_{\!\mu} T^{\mu\nu} = H^\nu_{\ \;\rho\sigma} J^{\rho\sigma}\,,\ \ \ \ \ \ 
  \nabla_{\!\mu} J^{\mu\nu} = 0\,.
\end{equation}
These are the same equations (\ref{eq:TC2}), (\ref{eq:BI}) we had earlier. The conserved external current is taken as $J^\mu_{\rm ext} = \coeff12 \epsilon^{\mu\nu\rho\sigma} \partial_{\nu} \Pi_{\rho\sigma}^{\rm ext}$, where $\Pi^{\rm ext}_{\mu\nu}$ may be viewed as the dual of the external polarization tensor~$M^{\mu\nu}_{\rm ext}$. The coupling $A_\mu J^\mu_{\rm ext}$ then becomes $\frac12 \Pi^{\rm ext}_{\mu\nu} J^{\mu\nu}$ upon integration by parts, and correlation functions of $J^{\mu\nu}$ may be obtained as variations with respect to $\Pi^{\rm ext}_{\mu\nu}$. The tensor $H$ in (\ref{eq:GHI-eom}) is $H=\frac12 d\Pi^{\rm ext}$, or in components $H_{\alpha\beta\gamma} = \frac14\partial_\alpha \Pi_{\beta\gamma}^{\rm ext}$ + (signed permutations).

In order to relate the GHI thermodynamic parameters to ours, we can compare equilibrium currents. The currents at zeroth order in derivatives are given by
\begin{subequations}
\label{eq:GHI-TJ0}
\begin{align}
  & T^{\mu\nu} = (\varepsilon_{\GHI} + p_{\GHI}) u^\mu u^\nu + p_{\GHI}\, g^{\mu\nu} - \mu_{\GHI}\, \rho_{\GHI}\, h^\mu h^\nu + O(\partial)\,,\\
  & J^{\mu\nu} = \rho_\GHI (u^\mu h^\nu - u^\nu h^\mu) +O(\partial)\,.
\end{align}
\end{subequations}
The subscript ``d'' for ``dual'' is used to differentiate the parameters from those used earlier in the paper. 
The currents can be compared with our eq.~(\ref{eq:TTF}) and the dual of eq.~(\ref{eq:FEB}) at zeroth order:
\begin{subequations}
\label{eq:TJ0-22}
\begin{align}
  & T^{\mu\nu} = \left( w_\m + \frac{B^2}{\mu_\m} \right)u^\mu u^\nu + \left( -\coeff12 B^2 + p_\m + \frac{B^2}{\mu_\m}\right) g^{\mu\nu} - \frac{B^\mu B^\nu}{\mu_\m} + O(\partial)\,,\\
  & J^{\mu\nu} = u^\mu B^\nu - u^\nu B^\mu  + O(\partial)\,,
\end{align}
\end{subequations}
where $w_\m \equiv T p_{\m,T} + \mu p_{\m,\mu} = Ts + \mu n$ is the enthalpy density, and $\mu_\m = 1/(1 - 2\partial p_\m/\partial B^2)$ is the magnetic permeability. Using $h^2=1$, we can identify $\rho_\GHI = B$, $\mu_\GHI = B/\mu_\m$, $h^\mu = B^\mu/B$, $p_\GHI = -\frac12 B^2 + p_\m + B^2/\mu_\m$, up to $O(\partial)$ terms. Out of equilibrium, $h^\mu$ and $\mu_\GHI$ are auxiliary dynamical variables (without a unique microscopic definition) designed to capture the dynamics of the magnetic field. The entropy density is $s_\GHI = p_{\m,T} + \frac{\mu}{T} p_{\m,\mu}$, as follows from $\varepsilon_\GHI + p_\GHI = T s_\GHI + \mu_\GHI \rho_\GHI$. The energy densities coincide, $\varepsilon_\GHI = -p + T s + \mu n = \epsilon$, again with $p=-\frac12 B^2 + p_\m(T,\mu,B^2)$.

At order $O(\partial)$, our constitutive relations can not be directly compared to those of GHI because of different hydrodynamic variables. However, we can compare the number of transport coefficients. 
The comparison may be done based on the entropy current argument which we review below.

In a particular hydrodynamic ``frame'', the one-derivative contributions to the GHI constitutive relations are given in eq.~(3.4), (3.5) of ref.~\cite{Grozdanov:2016tdf},
\begin{subequations}
\label{eq:GHI-TJ1}
\begin{align}
  & T^{\mu\nu}_{(1)} = \delta\! f_\GHI \, \Delta^{\mu\nu}_\GHI
    + \delta\tau_\GHI \, h^\mu h^\nu 
    + \ell_{\GHI}^\mu h^\nu + \ell_{\GHI}^\nu h^\mu 
    + t_\GHI^{\mu\nu}\,,\\
  & J^{\mu\nu}_{(1)} =  m_{\GHI}^\mu h^\nu - m_{\GHI}^\nu h^\mu 
    + s_\GHI^{\mu\nu}\,,
\end{align}
\end{subequations}
where $\Delta^{\mu\nu}_\GHI = g^{\mu\nu} + u^\mu u^\nu -  h^\mu h^\nu$, and 
the coefficients $\delta\! f_\GHI$, $\delta\tau_\GHI$, $\ell_{\GHI}^\mu$, $t_\GHI^{\mu\nu}$, $m_{\GHI}^\mu$, $s_\GHI^{\mu\nu}$ are all $O(\partial)$. 
The quantities $\ell_{\GHI}^\mu$, $t_\GHI^{\mu\nu}$, $m_{\GHI}^\mu$, $s_\GHI^{\mu\nu}$ are all transverse to both $u_\mu$ and $h_\mu$, the tensor $t_\GHI^{\mu\nu}$ is symmetric and traceless, and the tensor $s_\GHI^{\mu\nu}$ is anti-symmetric. We do not write the subscript on the temperature and fluid velocity, even though the GHI's $T$ and $u^\mu$ differ from ours at $O(\partial)$. 
Further, GHI impose charge conjugation as a constraint on the dynamics.

\subsection{Entropy production}
The ``canonical'' entropy current in the GHI formulation is analogous to eq.~(\ref{eq:S-canon}),
\begin{equation}
\label{eq:GHI-S-canon}
  S^\mu_\GHI = \frac{1}{T} \left( p_\GHI u^\mu - T^{\mu\nu} u_\nu  - \mu_\GHI J^{\mu\nu} h_\nu \right)\,.
\end{equation}
This does not take into account the $O(\partial)$ contributions to thermodynamics: as we have seen earlier, the only non-trivial thermodynamic susceptibility in a parity-invariant theory is odd under charge charge conjugation C, and gets eliminated if C is imposed as a symmetry of hydrodynamics.

Upon using the conservation equations (\ref{eq:GHI-eom}) together with the zeroth-order constitutive relations~(\ref{eq:GHI-TJ0}), the divergence of the entropy current (\ref{eq:GHI-S-canon}) is
$$
  \nabla_{\!\mu} S^\mu_\GHI = -T^{\mu\nu}_{(1)}\, \nabla_{\!\mu}\left(\frac{u_\nu}{T}\right)
  - J^{\mu\nu}_{(1)} \left[ \nabla_{\!\mu} \left( \frac{\mu_\GHI h_\nu}{T}\right) + \frac{u_\alpha H^{\alpha}_{\ \;\mu\nu}}{T}\right]\,.
$$
Substituting the first-order constitutive relations (\ref{eq:GHI-TJ1}), we find
\begin{align}
  T \nabla_{\!\mu} S^\mu_\GHI  = 
    -\delta\!f_\GHI\, (S_3 - S_4) - \delta \tau_\GHI S_4
    - \ell^\mu_\GHI \Sigma_\mu - \coeff12 t_{\mu\nu}^\GHI \sigma_\perp^{\mu\nu} - m_\alpha^\GHI \, Y^\alpha
    - \coeff12 s^\GHI_{\rho\sigma} Z^{\rho\sigma} \,.
\label{eq:GHI-DS}
\end{align}
Using the notation similar to sec.~\ref{sec:entropy-production}, we have the scalars $S_3\equiv \nabla{\cdot}u$, $S_4 \equiv h^\mu h^\nu \nabla_{\!\mu}u_\nu$, as well as
$\sigma^{\mu\nu}_\perp \equiv \coeff12 \left(\Delta_\GHI^{\mu\alpha} \Delta_\GHI^{\nu\beta} + \Delta_\GHI^{\nu\alpha} \Delta_\GHI^{\mu\beta} - \Delta_\GHI^{\mu\nu} \Delta_\GHI^{\alpha\beta}\right) \sigma_{\alpha\beta}$ and $\Sigma^\mu \equiv \Delta_\GHI^{\mu\lambda}\sigma_{\lambda\rho}h^\rho$. We have further defined
\begin{align*}
  & Y^\lambda \equiv \Delta_\GHI^{\lambda\rho} \left[ T\partial_\rho(\mu_\GHI/T) + 2 u_\alpha H^\alpha_{\ \rho\sigma} h^\sigma - \mu_\GHI h^\alpha \nabla_{\!\alpha} h_\rho \right] \,,\\
  & Z^{\alpha\beta} \equiv \Delta_\GHI^{\alpha\rho} \Delta_\GHI^{\beta\sigma} \left[ \mu_\GHI (\nabla_{\!\rho} h_\sigma - \nabla_{\!\sigma} h_\rho)
    + 2 u_\alpha H^\alpha_{\ \; \rho\sigma} \right]\,.
\end{align*}
In order to ensure that the entropy production in eq.~(\ref{eq:GHI-DS}) is non-negative, GHI demand
\begin{equation}
\label{eq:GHI-trcoefs}
\begin{aligned}
  & \delta\!f_\GHI = -\zeta_\perp (S_3 - S_4)\,,\ \ \ \ \ \ 
    \delta\tau_\GHI = -2\zeta_\parallel S_4\,,\ \ \ \ \ \ 
    \ell^\mu_\GHI = -\eta_\parallel \Sigma^\mu\,,\\
  & t^{\mu\nu}_\GHI = -\eta_\perp \sigma^{\mu\nu}_\perp\,,\ \ \ \ \ \ 
    m^\alpha_\GHI = -r_\perp  Y^\alpha \,,\ \ \ \ \ \ 
    s^{\rho\sigma}_\GHI = -r_\parallel Z^{\rho\sigma}\,,
\end{aligned}
\end{equation}
with six non-negative coefficients $\zeta_\perp$, $\zeta_\parallel$, $\eta_\perp$, $\eta_\parallel$, $r_\perp$, $r_\parallel$. This clearly gives $\nabla_{\!\mu} S^\mu_\GHI \geqslant 0$.

Note however that while demanding eq.~(\ref{eq:GHI-trcoefs}) is sufficient to ensure non-negative entropy production, there are more ways besides eq.~(\ref{eq:GHI-trcoefs}) to make the right-hand side of eq.~(\ref{eq:GHI-DS}) non-negative. These other options will give rise to extra transport coefficients. Indeed, consider the following coefficients of the $O(\partial)$ constitutive relations:
\begin{subequations}
\label{eq:trcoefs-dual}
\begin{align}
  & \delta\!f_\GHI = -f_1 S_3 - f_2 S_4 \,,\\
  & \delta\tau_\GHI = -\tau_1 S_3 - \tau_2 S_4\,,\\
  & \ell^\mu_\GHI = -\eta_\parallel \Sigma^\mu - \tilde\eta_\parallel \tilde\Sigma^\mu\,,\\
  & t^{\mu\nu}_\GHI = -\eta_\perp \sigma^{\mu\nu}_\perp - \tilde\eta_\perp \tilde\sigma^{\mu\nu}_\perp\,,\\
  & m^\alpha_\GHI = -r_\perp  Y^\alpha - \tilde r_\perp \tilde Y^\alpha\,,\\
  & s^{\rho\sigma}_\GHI = -r_\parallel Z^{\rho\sigma} \,.
\end{align}
\end{subequations}
The tilded vectors are defined as $\tilde V^\mu = \epsilon^{\mu\nu\alpha\beta}u_\nu h_\alpha V_\beta$, and the tilded shear tensor is
\begin{align*}
   \tilde\sigma_\perp^{\mu\nu} \equiv \coeff12 \left( \epsilon^{\mu\lambda\alpha}_{\ \ \ \ \beta} u_\lambda h_\alpha \sigma_{\perp}^{\beta\nu} + \epsilon^{\nu\lambda\alpha}_{\ \ \ \ \beta} u_\lambda h_\alpha \sigma_{\perp}^{\beta\mu}\right)\,, 
\end{align*}
as in eq.~(\ref{eq:sigma-tilde}). The tensor $s^{\rho\sigma}_\GHI$ has only one degree of freedom, hence it contains only one transport coefficient. The divergence of the entropy current (\ref{eq:GHI-DS}) is then
\begin{align}
   T \nabla_{\!\mu} S^\mu_\GHI  & =  
     f_1 S_3^2 +(\tau_1{+} f_2 {-} f_1 ) S_3 S_4 + (\tau_2 {-} f_2)S_4^2 \nonumber\\
   & + \eta_\parallel \Sigma_\mu \Sigma^\mu 
     + \coeff12 \eta_\perp (\sigma_\perp^{\mu\nu})^2 + r_\perp Y_\mu Y^\mu
     +\coeff12 r_\parallel (Z^{\rho\sigma})^2\,.
\label{eq:GHI-DS-2}
\end{align}
The three tilded coefficients do not contribute to entropy production in eq.~(\ref{eq:GHI-DS}) due to $\tilde V^\mu V_\mu = 0$ and $\sigma_{\!\perp\mu\nu}\,\tilde\sigma_{\perp}^{\mu\nu} = 0$, and can take any real values,
\begin{align}
  \tilde\eta_\parallel \in \mathbb{R}\,,\ \ \ \ \ \ 
  \tilde\eta_\perp \in \mathbb{R}\,,\ \ \ \ \ \ 
  \tilde r_\perp \in \mathbb{R}\,.
\end{align}
Demanding that $\nabla_{\!\mu} S^\mu_\GHI$ in eq.~(\ref{eq:GHI-DS-2}) is non-negative now implies
\begin{subequations}
\label{eq:constraints-dual}
\begin{align}
  \eta_\perp \geqslant 0\,,\ \ \ \ \ \ 
  \eta_\parallel \geqslant 0\,,\ \ \ \ \ \ 
  r_\perp \geqslant 0\,,\ \ \ \ \ \ 
  r_\parallel \geqslant 0\,,
\end{align}
together with the condition that the quadratic form in the first line of eq.~(\ref{eq:GHI-DS-2}) is positive semi-definite. The latter gives
\begin{align}
  f_1 \geqslant 0\,,\ \ \ \ \ \ 
  \tau_2 - f_2 \geqslant 0\,,\ \ \ \ \ \ 
  f_1 (\tau_2 - f_2) \geqslant \coeff14 (\tau_1 - f_1 + f_2)^2\,.
\end{align}
\end{subequations}
Thus there are eleven apriori independent non-equilibrium transport coefficients listed in Eqs.~(\ref{eq:trcoefs-dual}) that are consistent with non-negative entropy production, provided the constraints~(\ref{eq:constraints-dual}) are satisfied.
The coefficients $\tilde r_\perp$, $\tilde\eta_\perp$, $\tilde\eta_\parallel$ are odd under charge conjugation~C, and can be eliminated if one demands C-invariance of hydrodynamics. An implicit assumption of ref.~\cite{Grozdanov:2016tdf} amounts to choosing $f_1 = -f_2 = \zeta_\perp$, $\tau_1 = 0$, $\tau_2 = 2\zeta_\parallel$.

\subsection{Kubo formulas}
Assuming time-reversal covariance, the above transport coefficients can be further constrained by the Onsager relation~(\ref{eq:OR}). In order to find the retarded functions, we can use exactly the same variational procedure as in sec.~\ref{sec:correlators-2}:
\begin{subequations}
\label{eq:correlators-dual}
\begin{align}
  & G_{T^{\mu\nu} T^{\alpha\beta}} = \frac{2\,\delta}{\delta g_{\alpha\beta}} \left( \sqrt{-g}\, T^{\mu\nu}_\textrm{on-shell}[\Pi^{\rm ext},g] \right)\,,
  &  G_{J^{\mu\nu} T^{\alpha\beta}} = \frac{2\,\delta}{\delta g_{\alpha\beta}} \left( \sqrt{-g}\, J^{\mu\nu}_\textrm{on-shell}[\Pi^{\rm ext},g] \right)\,,
\end{align}
as well as
\begin{align}
  & G_{T^{\mu\nu} J^{\alpha\beta}} = 2\frac{\delta}{\delta \Pi_{\alpha\beta}^{\rm ext}}  T^{\mu\nu}_\textrm{on-shell}[\Pi^{\rm ext},g]  \,,
  &  G_{J^{\mu\nu} J^{\alpha\beta}} = 2\frac{\delta}{\delta \Pi_{\alpha\beta}^{\rm ext}}  J^{\mu\nu}_\textrm{on-shell}[\Pi^{\rm ext},g] \,.
\end{align}
\end{subequations}
Again, the subscript ``on-shell'' signifies that $T^{\mu\nu}$ and $J^{\mu\nu}$ are evaluated on the solutions to the conservation equations~(\ref{eq:GHI-eom}) with the constitutive relations (\ref{eq:trcoefs-dual}). We use the above prescription to evaluate correlation functions at zero spatial momentum, which gives rise to Kubo formulas. Demanding that the correlation functions satisfy (\ref{eq:OR}) now gives the Onsager relation
\begin{equation}
\label{eq:OR-dual}
  \tau_1 = f_1 + f_2\,.
\end{equation}
We further find the following Kubo formulas for transport coefficients in the constitutive relations~(\ref{eq:trcoefs-dual}). The resistivities are given by
\begin{subequations}
\begin{align}
  & \coeff{1}{\omega}{\rm Im}\, G_{J^{xy} J^{xy}}(\omega,{\bf k} {=}0) = r_\parallel\,,\\[5pt]
  & \coeff{1}{\omega}{\rm Im}\, G_{J^{xz} J^{xz}}(\omega,{\bf k} {=}0) = r_\perp\,,\\[5pt]
  & \coeff{1}{\omega}{\rm Im}\, G_{J^{yz} J^{xz}}(\omega,{\bf k} {=}0) = \tilde r_\perp\, {\rm sign}(B_0)\,,
\end{align}
the ``shear viscosities'' are given by
\begin{align}
  & \coeff{1}{\omega}{\rm Im}\, G_{T^{xz} T^{xz}}(\omega,{\bf k} {=}0) = \eta_\parallel\,,\ \ \ \ \ \ 
  &&  \coeff{1}{\omega}{\rm Im}\, G_{T^{xy} T^{xy}}(\omega,{\bf k} {=}0) = \eta_\perp\,,\\[5pt]
  & \coeff{1}{\omega}{\rm Im}\, G_{T^{yz} T^{xz}}(\omega,{\bf k} {=}0) = \tilde\eta_\parallel\, {\rm sign}(B_0)\,,\ \ \ \ \ \ 
  && \coeff{1}{\omega}{\rm Im}\, G_{T^{xy} T^{xx}}(\omega,{\bf k} {=}0) = \tilde\eta_\perp\, {\rm sign}(B_0)\,,
\end{align}
and the ``bulk viscosities'' are given by
\begin{align}
\label{eq:TxxTxx-dual}
  & \coeff{1}{\omega}{\rm Im}\, G_{T^{xx} T^{xx}}(\omega,{\bf k} {=}0) = f_1 + \eta_\perp\,,\\[5pt]
  & \coeff{1}{\omega}{\rm Im}\, G_{T^{xx} T^{zz}}(\omega,{\bf k} {=}0) = f_1 + f_2\,,\\[5pt]
  & \coeff{1}{\omega}{\rm Im}\, G_{T^{zz} T^{zz}}(\omega,{\bf k} {=}0) = \tau_1 + \tau_2\,.
\end{align}
\end{subequations}
Correlation functions at non-zero momentum may also be found by using the above variational procedure.

\subsection{Mapping of transport coefficients}
We can compare the correlation functions of $T^{\mu\nu}$ and $J^{\mu\nu}$ evaluated using  (\ref{eq:correlators-dual}) with the correlation functions found in sec.~\ref{sec:correlators-2}. If the two approaches to MHD (section~\ref{sec:hydro-2} and section~\ref{sec:comparison-GHI}) compute the same physical objects $G_{T^{\mu\nu} T^{\alpha\beta}}$ etc, the results should agree. Comparing correlation functions at zero spatial momentum allows one to relate the transport coefficients in the constitutive relations~(\ref{eq:trcoefs-dual}) to transport coefficients introduced in section~\ref{sec:hydro-1}, see eq.~(\ref{eq:TTF}), (\ref{eq:JTF}). Doing so in the (dynamically) neutral state with $n_0=0$ gives the following relations. The resistivities are related by
\begin{subequations}
\label{eq:trcoefs-map}
\begin{align}
  r_\parallel = \frac{1}{\sigma_\parallel}\,,\ \ \ \ \ \ 
  r_\perp = \frac{\sigma_\perp}{\sigma_\perp^2 + \tilde\sigma^2}\,,\ \ \ \ \ \ 
  \tilde r_\perp = -\frac{\tilde\sigma}{\sigma_\perp^2 + \tilde\sigma^2}\,,
\end{align}
the ``shear viscosities'' $\eta_\perp$, $\tilde\eta_\perp$, $\eta_\parallel$, $\tilde\eta_\parallel$ agree, and the ``bulk viscosities'' are related by
\begin{align}
  & f_1 = \zeta_1 -\coeff23 \eta_1 \,, 
  && f_2 = \zeta_2 - \coeff23 \eta_2 \,,\\
  & \tau_1 = \zeta_1 +\coeff43 \eta_1 \,,
  && \tau_2 = \zeta_2 + \coeff43 \eta_2 \,.
\end{align}
\end{subequations}
The Onsager relation (\ref{eq:OR2}) maps to the Onsager relation~(\ref{eq:OR-dual}), as expected. The entropy current constraints (\ref{eq:entropy-constraints}) map to the entropy current constraints (\ref{eq:constraints-dual}), as expected. 

Finally, the mapping of transport coefficients (\ref{eq:trcoefs-map}) can be used to compare the eigenfrequencies of small oscillations of the (dynamically) neutral state found in eq.~(\ref{eq:Alfven-waves}), (\ref{eq:ms-waves}) to those found in ref.~\cite{Grozdanov:2016tdf}. Using the map of thermodynamic parameters spelled out below eq.~(\ref{eq:TJ0-22}), the speed of Alfv\'en waves agrees with ref.~\cite{Grozdanov:2016tdf}. The damping coefficient of Alfv\'en waves in eq.~(\ref{eq:Alfven-waves}) agrees with ref.~\cite{Grozdanov:2016tdf} when $B^2/\mu_\m \ll \epsilon+p$. The speed of magnetosonic waves in eq.~(\ref{eq:ms-speed}) agrees with ref.~\cite{Grozdanov:2016tdf}: in order to see this, note that the assumption of constant magnetic permeability amounts to assuming that the equation of state takes the form $p_\GHI = \frac12 \mu_\m \mu_\GHI^2 + F(T)$, or $p=-\frac{1}{2\mu_\m}B^2 + F(T)$, with some $F(T)$. In general, the speed of magnetosonic waves derived from the formalisms of sec.~\ref{sec:hydro-2} and sec.~\ref{sec:comparison-GHI} will not agree, except when $B^2/\mu_\m \ll (\epsilon+p)$. One reason is that the chemical potential for the electric charge is treated as a thermodynamic variable in sec.~\ref{sec:hydro-2}, hence the magnetosonic wave speed will in general depend on the charge susceptibility $(\partial n/\partial\mu)_{\mu=0}$. This thermodynamic derivative is not present in the formalism of sec.~\ref{sec:comparison-GHI}. Finally, note that the transport coefficient $\tau_1$ contributes to damping of fast magnetosonic waves, for example at $\theta=0$ we have $\Gamma_{\rm ms} = (\tau_1 + \tau_2)/(Ts_\GHI)$, in agreement with eq.~(\ref{eq:Gms-fast-theta0}).

\section{Discussion}

In this paper we have presented the equations of relativistic magnetohydrodynamics, by which we mean the hydrodynamics of a conducting fluid in local thermal equilibrium, with dynamical electromagnetic fields. MHD is naturally formulated in a derivative expansion with magnetic field $B\sim O(1)$. Electric screening does not imply that the electric field vanishes: rather, it implies $E\sim O(\partial)$ is subleading in the derivative expansion. We have adopted the simplest ``mean-field'' formulation in which the constitutive relations in the theory with dynamical electromagnetic fields are inherited from the theory with external electromagnetic fields. Our main focus was on transport coefficients. For a parity-symmetric microscopic system, we find eleven transport coefficients at one-derivative order. One transport coefficient is thermodynamic: it is a part of the equation of state in curved space, and contributes to flat-space correlations. Transport coefficients of this type in relativistic hydrodynamics were first identified in \cite{Baier:2007ix} where they appeared at second order in derivatives. In 2+1 dimensional hydrodynamics, thermodynamic transport coefficients can already appear at first order in derivatives~\cite{Jensen:2011xb}. Of the remaining ten transport coefficients, three are non-equilibrium and non-dissipative, and seven are non-equilibrium and dissipative. There are more transport coefficients for parity-violating fluids, as listed in sec.~\ref{sec:hydro-1}. We now comment on questions not discussed in detail in the main body of the paper.

\begin{itemize}
\item[]{\it Angular momentum generated by the magnetic field.---} The thermodynamic transport coefficient $\MO$ determines the response of equilibrium magnetic polarization to vorticity, as can be seen from eq.~(\ref{eq:m-vector}). One way to view $\MO$ is to note that a system of charged particles in external magnetic field will develop angular momentum. One can see this in the thermodynamic framework of sec.~\ref{sec:thermodynamics}. For a bounded system, the equilibrium energy-momentum tensor obtained by varying the equilibrium free energy (\ref{eq:Ws}), (\ref{eq:F1}) with respect to the metric will have a boundary contribution after the variation $\MO B {\cdot} \delta_g \Omega$ is integrated by parts~\cite{Kovtun:2016lfw}. The surface momentum density ${\cal Q}^\alpha_{\rm s} = \MO \epsilon^{\alpha\mu\nu\rho} u_\mu B_\nu n_\rho$ (where $n^\mu$ is the unit spacelike normal vector to the boundary) will give rise to angular momentum induced by the magnetic field. 
Consider a system at rest in flat space at constant temperature, charge density, and constant magnetic field ${\bf B}$. The angular momentum ${\bf L}$ derived from the energy-momentum tensor only receives a boundary contribution,
and one finds
$$
  \frac{{\bf L}}{V} = 2\MO {\bf B}\,,
$$
where $V$ is the spatial volume. In this sense $\MO$ determines ``angular momentum density''. As the coefficient $\MO$ is odd under charge conjugation~C, this generation of angular momentum only happens in a C-invariant theory if the equilibrium state has non-zero charge density. Similarly, for a system not subject to the magnetic field, in flat space, which rotates uniformly with small (namely $|{\bm \omega}|R\ll 1$ where $R$ is the size of the system) angular velocity ${\bm\omega}$, the magnetization density is ${\bf m} = 2\MO\, {\bm\omega}$. More generally, the susceptibility $\MO$ provides a macroscopic parametrization of gyromagnetic phenomena such as the Barnett and Einstein-de Haas effects.

\item[]{\it Previous work on transport coefficients.---} 
Papers \cite{Huang:2011dc, Finazzo:2016mhm} studied transport coefficients for relativistic fluids subject to an external magnetic field. While this does not correspond to MHD in the sense described in this paper (we define MHD as a theory in which magnetic field or its auxiliary is a dynamical degree of freedom), a fluid in external field is a fundamental building block for MHD. Parts of Refs.~\cite{Huang:2011dc, Finazzo:2016mhm} overlap with our Section~\ref{sec:hydro-1}.
Some of our results differ from those in Refs.~\cite{Huang:2011dc, Finazzo:2016mhm}: the analysis of thermodynamics, the number of transport coefficients, constraints on transport coefficients imposed by the positivity of entropy production, and some of the Kubo formulas. The details are given in Appendix~\ref{app:comparison}.

\item[]{\it Dual formulation of magneto-hydrodynamics.---} In sec.~\ref{sec:comparison-GHI} we compared our results with the recent ``dual'' formulation of MHD in ref.~\cite{Grozdanov:2016tdf}. We found the same number of transport coefficients in the two approaches,
provided the bulk viscosity missed in ref.~\cite{Grozdanov:2016tdf} is restored, and the constraint of C-invariance imposed in ref.~\cite{Grozdanov:2016tdf} is lifted.
It would be interesting to investigate the relation between the ``dual'' and ``conventional'' formulations of MHD further, in particular with regard to the description of electric charge fluctuations.

\item[]{\it Applicability regime.---} The MHD described in this paper treats electromagnetic fields classically. This means that the electromagnetic coupling constant must be small so that  quantum fluctuations of the electromagnetic field can be ignored. The applicability regime of MHD also includes $B\ll T^2$ (or restoring the fundamental constants $\hbar c e B \ll (k_{\rm B}T)^2$), as is necessary to restrict the hydrodynamic degrees of freedom to those inherited from thermodynamics. We do not have a method to systematically incorporate the effects of larger magnetic fields within the MHD description of sec.~\ref{sec:hydro-2}. The classical hydrodynamic theory also ignores statistical fluctuations, which are known to invalidate classical second-order hydrodynamics in 3+1 dimensions (and classical first-order hydrodynamics in 2+1 dimensions). Understanding the effects of statistical fluctuations in magnetic field requires further work. 

\item[]{\it Transport coefficients at strong coupling.---} While the small electromagnetic coupling allows one to treat magnetic fields classically, other interactions in the theory do not have to be small. For strongly interacting non-abelian gauge theories in external $U(1)$ magnetic field, methods of gauge-gravity duality provide a window into non-equilibrium physics, both within and outside the hydrodynamic regime. Some of the hydrodynamic transport coefficients discussed in this paper were evaluated in holographic models in refs.~\cite{Critelli:2014kra, Finazzo:2016mhm}. The full set of transport coefficients for fluids in external magnetic field has not yet been explored holographically. 

\item[]{\it Higher-order terms.---} We have not taken into account the terms beyond first order in the derivative expansion. In conventional hydrodynamics, higher-order terms are required to render the theory causal~\cite{Israel:1979wp} (see e.g.\ \cite{Baier:2007ix, Pu:2009fj} for more recent discussions). We expect that a causal formulation of MHD will involve higher-order relaxation times as well as the electric field dynamics.

\end{itemize}

\noindent
{\bf Note added:} We have communicated with the authors of ref.~\cite{Grozdanov:2016tdf}, and it is our understanding that the missing bulk viscosity will be added in an updated version of ref.~\cite{Grozdanov:2016tdf}, and that the Kubo formulas for bulk viscosities will agree with ours. We have also communicated with the authors of ref.~\cite{Finazzo:2016mhm}, and it is our understanding that the Kubo formulas for viscosities in an updated version of~ref.~\cite{Finazzo:2016mhm} will agree with ours. 

\acknowledgments
We thank the Perimeter Institute for Theoretical Physics where a large part of this work was completed. We thank the authors of refs.~\cite{Grozdanov:2016tdf,Finazzo:2016mhm} for discussing their papers and the connections to our work. We thank Shira Chapman, Kristan Jensen, and Adam Ritz for helpful conversations. This work was supported in part by NSERC of Canada.

\appendix

\section{Equilibrium $T^{\mu\nu}$ and $J^\mu$}
\label{app:thermo}
The coefficients $\epsilon_n$, $\pi_n$, $\phi_n$, $\gamma_n$, $\delta_n$, $\theta_n$ in the equilibrium energy-momentum tensor and the current (\ref{eq:TJeq}) have the following expressions in terms of the five parameters $M_n(T,\mu,B^2)$ of the generating functional (\ref{eq:F1}). The $O(\partial)$ correction to the energy density is determined by
\begin{align*}
  & \epsilon_1 = -M_1 + T M_{1,T} + \mu M_{1,\mu} + 4B^2 M_{1,B^2} + T^4 M_{3,B^2} \,,\\
  & \epsilon_2 = -M_2 + T M_{2,T} + \mu M_{2,\mu}\,,\\
  & \epsilon_3 = \frac{4B^2}{T^4} \left( M_1 - T M_{1,T} - \mu M_{1,\mu} - 4B^2 M_{1,B^2} \right) - 4B^2 M_{3,B^2}\,, \\
  & \epsilon_4 = -2M_4 + T M_{4,T} + \mu M_{4,\mu}\,, \\
  & \epsilon_5 = T M_{5,T} + \mu M_{5,\mu} + \frac{4B^2}{T^4} M_{1,\mu} + M_{3,\mu} \,,
\end{align*}
where the comma denotes the partial derivative: $M_{1,T} \equiv (\partial M_1/\partial T)$ evaluated at fixed $\mu$ and $B^2$, etc. The $O(\partial)$ correction to the pressure is determined by
\begin{align*}
  & \pi_1 = 0\,,\\
  & \pi_2 = -\coeff23 M_2 - \coeff43 B^2 M_{2,B^2} \,, \\
  & \pi_3 = -\coeff43 B^2 M_{3,B^2} + \frac{4B^2}{3T^4} \left(M_1 - T M_{1,T} - \mu M_{1,\mu} - 4B^2 M_{1,B^2} \right) \,, \\
  & \pi_4 = -\coeff13 M_4 - \coeff43 B^2 M_{4,B^2} \,,\\
  & \pi_5 = -\coeff43 B^2 M_{5,B^2} + \frac{4B^2}{3T^4}M_{1,\mu}\,.
\end{align*}
The $O(\partial)$ correction to the charge density is determined by
\begin{align*}
  & \phi_1 = M_{1,\mu} -T^4 M_{5,B^2} \,,\\
  & \phi_2 = M_{2,\mu} \,,\\
  & \phi_3 = M_{3,\mu} + T M_{5,T} + \mu M_{5,\mu} + 4B^2 M_{5,B^2} \,,\\
  & \phi_4 = -\aBB + M_{4,\mu} \,,\\
  & \phi_5 = 0\,. 
\end{align*}
The $O(\partial)$ correction to the energy flux is determined by
\begin{align*}
  & \gamma_1 = -M_4 \,,\\
  & \gamma_2 = 2M_4 - T M_{4,T} - \mu M_{4,\mu}  \,,\\
  & \gamma_3 = -M_{4,B^2} \,,\\
  & \gamma_4 = -\aBB + M_{4,\mu} \,.
\end{align*}
The $O(\partial)$ correction to the spatial current is determined by the magnetic susceptibility,
\begin{align*}
  & \delta_1 = -\aBB \,,\\
  & \delta_2 = \aBB - T \aBB{}_{,T} - \mu \aBB{}_{,\mu}    \,,\\
  & \delta_3 = -\aBB{}_{,B^2}  \,,\\
  & \delta_4 =  \aBB{}_{,\mu} \,.
\end{align*}
The $O(\partial)$ correction to the stress is determined by
\begin{align*}
  & \theta_1 = 0  \,,\\
  & \theta_2 = M_{2,B^2} \,,\\
  & \theta_3 = M_{3,B^2} -\frac{1}{T^4}\left(M_1 - T M_{1,T} -\mu M_{1,\mu} -4B^2 M_{1,B^2} \right) \,,\\
  & \theta_4 = M_{4,B^2} \,,\\
  & \theta_5 = M_{5,B^2} - \frac{1}{T^4} M_{1,\mu}\,, \\
  & \theta_6 = 2M_2\,,\\
  & \theta_7 = -M_2 + T M_{2,T} + \mu M_{2,\mu} \,, \\
  & \theta_8 = M_{2,B^2}\,, \\
  & \theta_9 = -M_{2,\mu}\,,\\
  & \theta_{10} = M_4\,.
\end{align*}

\section{Comparison with previous work}
\label{app:comparison}
\subsection{Comparison with Huang et al}
\label{app:comparison-HSR}
In this appendix we will comment on how our work relates to some earlier studies of transport coefficients, for the benefit of the reader who might want to compare different approaches. Ref.~\cite{Huang:2011dc}, abbreviated below as HSR, studied relativistic hydrodynamics of parity-invariant fluids in external non-dynamical magnetic field. HSR enumerated the transport coefficients, giving a relativistic version of the classification in the book~\cite{LL10}, \S 13, and derived the Kubo formulas for transport coefficients in an operator formalism. Parts of the HSR paper overlap with our Section~\ref{sec:hydro-1}.

Our counting of non-equilibrium transport coefficients for parity-invariant systems agrees with HSR. Denoting the transport coefficients in ref.~\cite{Huang:2011dc} with the subscript HSR, the relations to our transport coefficients are as follows:
\begin{equation}
\label{eq:HSR-coefs}
\begin{aligned}
  & \eta_\perp = \eta_{0, \HSR}\,,\ \ \ \ \ \ 
    \tilde\eta_\perp = -2\eta_{3, \HSR}\,,\ \ \ \ \ \ 
    \eta_\parallel = \eta_{0,\HSR} + \eta_{2, \HSR}\,,\ \ \ \ \ \ 
    \tilde\eta_\parallel = -\eta_{4,\HSR}\,,\\
  & \eta_1 = -\coeff12 \eta_{0,\HSR} -\coeff38 \eta_{1,\HSR} - \coeff34 \zeta_{\perp,\HSR}\,,\ \ \ \ \ \ \zeta_1 = \zeta_{\perp, \HSR}\,,\\
  &  \eta_2 = \coeff32 \eta_{0,\HSR} + \coeff98 \eta_{1,\HSR} + \coeff34 \zeta_{\perp,\HSR} + \coeff32 \zeta_{\parallel,\HSR} \,,\ \ \ \ \ \ \zeta_2 = \zeta_{\parallel, \HSR} -  \zeta_{\perp, \HSR}\,,\\
  & \sigma_\perp = \kappa_{\perp, \HSR}\,,\ \ \ \ \ \ 
    \sigma_\parallel = \kappa_{\parallel, \HSR}\,,\ \ \ \ \ \ 
    \tilde\sigma = -\kappa_{\times, \HSR}\,,
\end{aligned}
\end{equation}
assuming the convention $\epsilon^{0123}=1$.
This lists eleven transport coefficients compared to ten HSR coefficients, hence under this mapping the eleven transport coefficients are not independent. Indeed, the comparison (\ref{eq:HSR-coefs}) implies $\zeta_2 = 2\eta_1 + \coeff23 \eta_2$, which is precisely our Onsager constraint~(\ref{eq:OR2}). Thus our counting of non-equilibrium transport coefficients in Section~\ref{sec:hydro-1} agrees with that of HSR. 

There are also some differences between our Section~\ref{sec:hydro-1} and HSR. In terms of the setup, the HSR treatment neglects electric fields, while we include them and explain how to do so systematically. Related to that, the treatment of polarization effects in HSR was incomplete. A direct way to obtain the equilibrium energy-momentum tensor and the current in the presence of external fields is by varying the corresponding generating functional with respect to the metric and the gauge field, as was done for example in ref.~\cite{Kovtun:2016lfw}. As a result, HSR did not include the thermodynamic transport coefficient, denoted in Section~\ref{sec:hydro-1} as $\MO$, and did not distinguish between the Landau-Lifshitz and thermodynamic frames. In the Landau-Lifshitz frame, $\MO$ would contribute to all frame invariants in eq.~(\ref{eq:fi-2}) inducing $O(\partial)$ contributions to pressure, electric current, and spatial stress.

We also find that our constraints on transport coefficients imposed by the positivity of entropy production differ somewhat from those presented in HSR. Rewriting our constraints (\ref{eq:entropy-constraints}) in terms of the HSR coefficients, we find
\begin{equation}
\label{eq:HSR-constraints}
\begin{aligned}
 & \eta_{0, \HSR} \geqslant 0\,,\ \ \ \ \ \ 
   \eta_{0, \HSR} + \eta_{2, \HSR} \geqslant 0\,,\ \ \ \ \ \ 
   \coeff13 \eta_{0, \HSR} + \coeff14 \eta_{1, \HSR}
   + \coeff32 \zeta_{\perp, \HSR} \geqslant 0\,,\\
 & 3\eta_{0, \HSR} + \coeff94 \eta_{1, \HSR}
   + \coeff32 \zeta_{\perp, \HSR} + 3\zeta_{\parallel, \HSR} \geqslant 0\,,\\
 & 18 \zeta_{\parallel, \HSR}\zeta_{\perp, \HSR} +4 \zeta_{\parallel, \HSR}\eta_{0, \HSR} + 3 \zeta_{\parallel, \HSR} \eta_{1, \HSR}
   + 8 \zeta_{\perp, \HSR}\eta_{0, \HSR} + 6 \zeta_{\perp, \HSR} \eta_{1, \HSR} \geqslant 0\,,\\
 & \kappa_{\perp, \HSR} \geqslant0\,,\ \ \ \ \ \ 
   \kappa_{\parallel, \HSR} \geqslant 0\,.
\end{aligned}
\end{equation}
On the other hand, the constraints coming from the second law in ref.~\cite{Huang:2011dc} state that all the dissipative HSR transport coefficients must be positive. We find that the constraints on dissipative transport coefficients~(\ref{eq:HSR-constraints}) are in fact weaker. In other words, the constraints of ref.~\cite{Huang:2011dc} are too restrictive: some of the dissipative transport coefficients in the HSR notation can be negative, while still satisfying (\ref{eq:HSR-constraints}), and therefore still leading to positive entropy production. 

Finally, there are differences between our Kubo formulas and those of HSR. In particular our Kubo formulas for conductivities transverse to the external magnetic field are markedly different. Comparing the correlation functions in the neutral state ($n_0=0$), the HSR Kubo formulas give the conductivities $\kappa_{\perp, \HSR}$ and $\kappa_{\times, \HSR}$ in terms of the $i\omega$ coefficient of the retarded current-current correlation functions at zero momentum. On the other hand, our Kubo formulas (\ref{eq:Kubo-r1b}), (\ref{eq:Kubo-r1c}) show that the coefficient of $i\omega$ vanishes, while the subleading coefficient in the small-$\omega$ expansion is determined by the resistivity rather than the conductivity. In the charged state, the term $n_0/B_0$ in our eq.~(\ref{eq:Kubo-r1c}) describes the standard Hall effect in the plane transverse to the magnetic field. The Hall effect appears to be missing from correlation functions in ref.~\cite{Huang:2011dc}.

\subsection{Comparison with Finazzo et al}
\label{app:comparison-FCRN}
In ref.~\cite{Finazzo:2016mhm} (abbreviated below as FCRN), the authors considered hydrodynamics with fixed non-dynamical magnetic field, and derived Kubo formulas for transport coefficients that appear in the energy-momentum tensor in the Landau-Lifshitz frame. FCRN use a variational approach to find the retarded functions of the energy-momentum tensor, and Appendix~B of FCRN overlaps with our Section~\ref{sec:hydro-1}. FCRN follow ref.~\cite{Huang:2011dc} in their constitutive relations for the energy-momentum tensor, so the comments in Section~\ref{app:comparison-HSR} apply to FCRN as well, where FCRN agree with ref.~\cite{Huang:2011dc}. In particular, FCRN did not include the thermodynamic transport coefficient $\MO$ that appears in the equilibrium free energy at one-derivative order.

FCRN use mostly the same convention for transport coefficients as HSR: $\eta_{0,\FCRN} = \eta_{0,\HSR}$, $\eta_{1,\FCRN} = \eta_{1,\HSR}$, $\eta_{4,\FCRN} = \eta_{4,\HSR}$, $\zeta_{\perp, \FCRN} = \zeta_{\perp, \HSR}$, $\zeta_{\parallel, \FCRN} = \zeta_{\parallel, \HSR}$, while $\eta_{2,\FCRN} = - \eta_{2,\HSR}$, $\eta_{3,\FCRN} = -2\eta_{3,\HSR}$, assuming the convention $\epsilon^{0123}=1$. The translation to our convention for transport coefficients can be done through eq.~(\ref{eq:HSR-coefs}). The convention for the variational retarded correlation functions used by FCRN differs from ours by an overall minus sign. 

We agree with FCRN's Kubo formulas for $\eta_{0,\FCRN}$, $\zeta_{\perp, \FCRN}$, and $\zeta_{\parallel, \FCRN}$. Our Kubo formulas for $\eta_{2,\FCRN}$ and $\eta_{3,\FCRN}$ differ from those in ref.~\cite{Finazzo:2016mhm} by a minus sign. Our Kubo formula for $\eta_{4,\FCRN}$ differs from that in ref.~\cite{Finazzo:2016mhm} by a factor of $1/4$. 
Our Kubo formula for $\eta_{1,\FCRN} + \coeff43 \eta_{0,\FCRN}$ differs from that in ref.~\cite{Finazzo:2016mhm} by a factor of $2$. 
Ref.~\cite{Finazzo:2016mhm} does not derive Kubo formulas for electrical conductivities in external magnetic field, so we can not compare those.

\bibliographystyle{JHEP}
\bibliography{mhd2}

\end{document}